\documentclass[preprint]{aastex}
\def\alf{Alfv\'en }
\def\detg{{\sqrt{-g}}}
\def\bnabla{{\bf \nabla}}
\def\bB{{\bf B}}
\def\bP{{\bf P}}
\def\bU{{\bf U}}
\def\bF{{\bf F}}
\def\bva{{\bf v}_{\rm A}}
\def\kdv{({\bf k} \cdot {\bf v}_{\rm A})}

\def\va{{\bf v}_{\rm A}}
\def\bk{{\bf k}}

\def\del{{\partial}}

\def\eps{\epsilon}

\def\sE{{\mathcal{E}}}
\def\order{{\mathcal{O}}}
\def\Lone{{\mathcal{L}}_1} 
\shortauthors{Gammie et al.}
\shorttitle{Numerical GRMHD}
\begin{document}
\title{HARM: A Numerical Scheme for General Relativistic Magnetohydrodynamics}

\author{Charles F. Gammie \altaffilmark{1,2,3} and 
	Jonathan C. McKinney\altaffilmark{2}} 

\affil{Center for Theoretical Astrophysics, University of Illinois \\
1002 W. Green St., Urbana, IL 61801, USA; gammie@uiuc.edu,
jcmcknny@uiuc.edu}

\author{G\'abor T\'oth}

\affil{Department of Atomic Physics, Lor\'and E\"otv\"os University \\
P\'azm\'any P\'eter s\'et\'any 1/A Budapest, H-1117, Hungary; 
gtoth@hermes.elte.hu}

\altaffiltext{1}{Department of Astronomy}
\altaffiltext{2}{Department of Physics}
\altaffiltext{3}{National Center for Supercomputing Applications}

\begin{abstract}

We describe a conservative, shock-capturing scheme for evolving the
equations of general relativistic magnetohydrodynamics.  The fluxes are
calculated using the Harten, Lax, and van Leer scheme.  A variant of
constrained transport, proposed earlier by T\'oth, is used to maintain a
divergence free magnetic field.  Only the covariant form of the metric
in a coordinate basis is required to specify the geometry.  We describe
code performance on a full suite of test problems in both special and
general relativity.  On smooth flows we show that it converges at second
order.  We conclude by showing some results from the evolution of a
magnetized torus near a rotating black hole.

\end{abstract}

\keywords{accretion, accretion disks, black hole physics,
Magnetohydrodynamics: MHD, Methods: Numerical}

\section{Introduction}

Quasars, active galactic nuclei (AGN), X-ray binaries, gamma-ray bursts,
and core-collapse supernovae are all likely powered by a central engine
subject to strong gravity, strong electromagnetic fields, and rotation.
For convenience, we will refer to this class of objects as {\it
relativistic magneto-rotators} (RMRs).  RMRs are among the most luminous
objects in the universe and are therefore the center of considerable
theoretical attention.  Unfortunately the governing physical laws for
RMRs, while well known, are nonlinear, time-dependent, and intrinsically
multidimensional.  This has stymied development of a first-principles
theory for their evolution and observational appearance and strongly
motivates a numerical approach.

To fully understand RMR structure, one must be able to follow the
interaction of a non-Maxwellian plasma with a relativistic gravitational
field, a strong electromagnetic field, and possibly a strong radiation
field as well.  This general problem remains beyond the reach of today's
algorithms and computers.  A useful first step, however, might be to
study these objects in a nonradiating magnetohydrodynamic (MHD) model.
In this case the plasma can be treated as a fluid, greatly reducing the
number of degrees of freedom, and the radiation field can be ignored.
The relevance of this approximation must be evaluated in astrophysical
context and will not be considered here.  

We were motivated, therefore, to develop a method for integrating the
equations of ideal, general relativistic MHD (GRMHD), and that method is
described in this paper.  This is in a sense well-trodden ground: many
schemes have already been developed for relativistic fluid dynamics.
What has not existed until recently is a scheme that: (1) includes
magnetic fields; (2) has been fully verified and convergence tested; (3)
is stable and capable of integrating a flow over many dynamical times.
A pioneering GRMHD code has been developed by Koide and collaborators
(e.g.  \cite{kskm}, \cite{ksk}, \cite{kmsk}, \cite{mku}).  Our code code
differs in that we have subjected it to a fuller series of tests
(described in this paper), we can perform longer integrations than the
rather brief simulations described in the published work of Koide's
group, and our code explicitly maintains the divergence-free constraint
on the magnetic field.  A GRMHD Godunov scheme based on a Roe-type
approximate Riemann solver has been developed by Komissarov and
described in a conference proceeding \citep{kom01}.

A rather complete review of numerical approaches to relativistic fluid
dynamics is given by \citet{mm99} and \citet{font00}.  The first
numerical GRMHD scheme that we are aware of is by \cite{wil77}, who
integrated the GRMHD equations in axisymmetry near a Kerr black hole.
While it was recognized throughout the 1970s that relativistic MHD would
be relevant to problems related to black hole accretion (particularly
following the seminal work of \cite{bz77} and \cite{ph83}; see also
\cite{pun01}), no further work appeared until \cite{yoko93}.  The
discovery by \cite{bh91} that magnetic fields play a crucial role in
regulating accretion disk evolution (reviewed in \cite{bh98}) and the
absence of purely hydrodynamic means for driving accretion in disks
\citep{bh98} further motivated the development of relativistic MHD
schemes.  More recently there have been several efforts to develop GRMHD
codes, including the already-mentioned work by Koide and collaborators
and by Komissarov.  A ZEUS-like scheme for GRMHD has also been developed
and is described in a companion paper \citep{dvh}.

Special relativistic MHD (SRMHD) is the foundation for any GRMHD scheme,
although there are nontrivial problems in making the transition to full
general relativity.  SRMHD schemes have been developed by
\cite{van93,bal01,kku,kom99} and \cite{dzbl}.  We were particularly
influenced by the clear development of the fundamental equations in
\cite{kom99} for his Godunov SRMHD scheme based on a Roe-type
approximate Riemann solver, and by the work of \cite{dzb} and
\cite{dzbl} who chose to use the simple approximate Riemann solver of
\cite{hll83} in their special relativistic hydrodynamics and SRMHD
schemes, respectively.

Our numerical scheme is called HARM, for High Accuracy Relativistic
Magnetohydrodynamics.\footnote{also named in honor of R. H\"arm, who
with M. Schwarzschild was a pioneer of numerical astrophysics.} In the
next section we develop the basic equations in the form used for
numerical integration in HARM (\S 2).  In \S 3 we describe the basic
algorithm.  In \S 4 we describe the performance of the code on a series
of test problems.  In \S 5 we describe a sample evolution of a
magnetized torus near a rotating black hole.

\section{A GRMHD Primer}

The equations of general relativistic MHD are well known, but for
clarity we will develop them here in the same form used in numerical
integration.  Unless otherwise noted $c = 1$ and we follow the
notational conventions of \cite{mtw}, hereafter MTW.  The reader may
also find it useful to consult \cite{anile}.

The first governing equation describes the conservation of particle
number:
\begin{equation}
(n u^{\mu})_{;\mu} = 0.
\end{equation}
Here $n$ is the particle number density and $u^\mu$ is the
four-velocity.  For numerical purposes we rewrite this in a coordinate
basis, replacing $n$ with the ``rest-mass density'' $\rho = m n$, where
$m$ is the mean rest-mass per particle:
\begin{equation}\label{PARTNOCONS}
{1\over{\detg}}\partial_\mu (\detg \,\rho u^\mu) = 0.
\end{equation}
Here $g \equiv Det(g_{\mu\nu})$.  

The next four equations express conservation of energy-momentum:
\begin{equation}
{T^{\mu}}_{\nu;\mu} = 0,
\end{equation}
where ${T^\mu}_\nu$ is the stress-energy tensor.  In a coordinate basis,
\begin{equation}\label{ENER_MOM}
\partial_t \left( \detg \, {T^t}_\nu \right)
	= -\partial_i \left( \detg \, {T^i}_\nu
	\right) + \detg \, {T^\kappa}_\lambda \, {\Gamma^\lambda}_{\nu \kappa},
\end{equation}
where $i$ denotes a spatial index and ${\Gamma^\lambda}_{\nu \kappa}$ is
the connection.  

The energy-momentum equations have been written with the free index down
for a reason.  Symmetries of the metric give rise to conserved currents.
In the Kerr metric, for example, the axisymmetry and stationary nature
of the metric give rise to conserved angular momentum and energy
currents.  In general, for metrics with an ignorable coordinate $x^\mu$
the source term on the right hand side of the evolution equation for
$T^t_\mu$ vanish.  These source terms do not vanish when the equation is
written with both indices up.

The stress-energy tensor for a system containing only a perfect fluid
and an electromagnetic field is the sum of a fluid part,
\begin{equation}
T^{\mu\nu}_{\rm fluid} = (\rho + u + p) u^\mu u^\nu + p g^{\mu\nu},
\end{equation}
(here $u \equiv$ internal energy and $p \equiv$ pressure)
and an electromagnetic part,
\begin{equation}
T^{\mu\nu}_{\rm EM} = F^{\mu\alpha} F^\nu_\alpha - {1\over{4}}
	g^{\mu\nu} F_{\alpha\beta} F^{\alpha\beta}.
\end{equation}
Here $F^{\mu\nu}$ is the electromagnetic field tensor (MTW:
``Faraday''), and for convenience we have absorbed a factor of
$\sqrt{4\pi}$ into the definition of $F$.

The electromagnetic portion of the stress-energy tensor simplifies if we
adopt the ideal MHD approximation, in which the electric field vanishes
in the fluid rest frame due to the high conductivity of the plasma
(``${\bf E} + {\bf v}\times{\bB} = 0$'').  Equivalently the Lorentz
force on a charged particle vanishes in the fluid frame:
\begin{equation}\label{IDEALMHD}
u_\mu F^{\mu\nu} = 0.
\end{equation}
It is convenient to define the magnetic field four-vector
\begin{equation}\label{B4DEF}
b^\mu \equiv {1\over{2}} \eps^{\mu\nu\kappa\lambda} u_\nu F_{\lambda\kappa},
\end{equation}
where $\eps$ is the Levi-Civita {\em tensor}.  Recall that (following
the notation of MTW) $\eps^{\mu\nu\lambda\delta} = -{1\over{\detg}}
[\mu\nu\lambda\delta]$, where $[\mu\nu\lambda\delta] $ is the completely
antisymmetric symbol and $= 0,1,$ or $-1$.  These can be combined (with
the aid of identity 3.50h of MTW):
\begin{equation}\label{FDEF}
F^{\mu\nu} = \eps^{\mu\nu\kappa\lambda} u_\kappa b_\lambda.
\end{equation}
Substitution and some manipulation (using identities 3.50 of MTW and
$b_\mu u^\mu = 0$; the latter follows from the definition of $b_\mu$ and
the antisymmetry of $F$) yields
\begin{equation}
T^{\mu\nu}_{\rm EM} = b^2 u^\mu u^\nu + {1\over{2}} b^2 g^{\mu\nu} - b^\mu b^\nu.
\end{equation}
Notice that the last two terms are nearly identical to the
nonrelativistic MHD stress tensor, while the first term is higher order
in $v/c$.  To sum up, 
\begin{equation}\label{TMUNUMHD}
T^{\mu\nu}_{\rm MHD} = (\rho + u + p + b^2) u^\mu u^\nu 
+ (p + {1\over{2}} b^2) g^{\mu\nu} - b^\mu b^\nu
\end{equation}
is the MHD stress-energy tensor.

The electromagnetic field evolution is given by the source-free part
of Maxwell's equations
\begin{equation}\label{MAXA}
F_{\mu\nu,\lambda} + F_{\lambda\mu,\nu} + F_{\nu\lambda,\mu} = 0.
\end{equation}
The rest of Maxwell's equations determine the current
\begin{equation}
J^\mu = {F^{\mu\nu}}_{;\nu},
\end{equation}
and are not needed for the evolution, as in nonrelativistic MHD.  

Maxwell's equations can be written in conservative form by taking
the dual of eq.(\ref{MAXA}):
\begin{equation}\label{REDMAX}
{F^{*\mu\nu}}_{;\nu} = 0.
\end{equation}
Here $F^*_{\mu\nu} = {1\over{2}}\epsilon_{\mu\nu\kappa\lambda}
F^{\kappa\lambda}$ is the dual of the electromagnetic field tensor (MTW:
``Maxwell'').  In ideal MHD 
\begin{equation}
F^{*\mu\nu} = b^\mu u^\nu - b^\nu u^\mu,
\end{equation}
which can be proved by taking the dual of eq.(\ref{FDEF}).  

The components of $b^\mu$ are not independent, since $b^\mu u_\mu = 0$.
Following, e.g., \cite{kom99}, it is useful to define the magnetic field
three-vector $B^i = F^{*it}$.  In terms of $B^i$,
\begin{equation}\label{BTSOL}
b^t = B^i u^\mu g_{i\mu},
\end{equation}
\begin{equation}\label{BISOL}
b^i = (B^i + b^t u^i)/u^t.
\end{equation}
The space components of the induction equation then reduce to
\begin{equation}\label{INDUCTION}
\del_t(\detg B^i) = -\del_j (\detg\, (b^j u^i - b^i u^j))
\end{equation}
and the time component reduces to 
\begin{equation}\label{DIVB}
{1\over{\detg}} \del_i (\detg\, B^i) = 0,
\end{equation}
which is the no-monopoles constraint.  The appearance of $B^i$ in these
last two equations are what motivates the introduction of the field
three-vector in the first place.

To sum up, the fundamental equations as used in HARM are: the particle
number conservation equation (\ref{PARTNOCONS}); the four
energy-momentum equations (\ref{ENER_MOM}), written in a coordinate
basis and using the MHD stress-energy tensor of equation
(\ref{TMUNUMHD}); and the induction equation (\ref{INDUCTION}), subject
to the constraint (\ref{DIVB}).  These hyperbolic \footnote{The GRMHD
equations exhibit the same degeneracies as the nonrelativistic MHD
equations.} equations are written in conservation form, and so can be
solved numerically by well-known techniques.

\section{Numerical Scheme}

There are many possible ways to numerically integrate the GRMHD
equations.  A first, zeroth-order choice is between conservative and
nonconservative schemes.  Nonconservative schemes such as ZEUS
\citep{sn92} have enjoyed wide use in numerical astrophysics.  They
permit the integration of an internal energy equation rather than a
total energy equation.  This can be advantageous in regions of a flow
where the internal energy is small compared to the total energy (highly
supersonic flows), which is a common situation in astrophysics.  A
nonconservative scheme for GRMHD following a ZEUS-like approach has been
developed and is described in a companion paper \citep{dvh}.

We have decided to write a conservative scheme.  One advantage of this
choice is that in one dimension, total variation stable schemes are
guaranteed to converge to a weak solution of the equations by the
Lax-Wendroff theorem \citep{lw60} and by a theorem due to \cite{lev98}.
While no such guarantee is available for multidimensional flows, this is
a reassuring starting point.  Furthermore, one is guaranteed that a
conservative scheme in any number of dimensions will satisfy the jump
conditions at discontinuities.  This is not true in artificial viscosity
based nonconservative schemes, which are also known to have trouble in
relativistic shocks \citep{nw86}.  Conservative schemes for GRMHD have
also been developed by \cite{kom01} and by \cite{ksk}.

A conservative scheme updates a set of ``conserved'' variables at each
timestep.  Our vector of conserved variables is
\begin{equation}
\bU \equiv \detg (\rho u^t, T^t_t, T^t_i, B^i).
\end{equation}
These are updated using fluxes $\bF$.  We must also choose a set of
``primitive'' variables, which are interpolated to model the flow within
zones.  We use variables with a simple physical interpretation: 
\begin{equation}
\bP = (\rho, u, v^i, B^i).
\end{equation}
Here $v^i = u^i/u^t$ is the 3-velocity.\footnote{We initially
used $u^i$ as primitive variables, but the inversion $u^t(u^i)$
is not always single-valued, e.g. inside the ergosphere of a
black hole.  That is, there are physical flows with the same
value of $u^i$ but different values of $u^t$.} The functions
$\bU(\bP)$ and $\bF(\bP)$ are analytic, but the inverse
operations (so far as we can determine) are not.
\footnote{\cite{dzbl} have found that the inversion $\bP(\bU)$
can be reduced analytically to the solution of a single,
nonlinear equation.} There is also no simple expression for
$\bF(\bU)$.  

To evaluate $\bU(\bP)$ and $\bF(\bP)$ one must find $u^t$ and
$b^\mu$ from $v^i$ and $B^i$.  To find $u^t$, solve the quadratic
equation $g_{\mu\nu} u^\mu u^\nu = -1$.  Next use equations
(\ref{BTSOL}) and (\ref{BISOL}) to find $b^\mu$; these require
only multiplications and additions. The remainder of the
calculation of $\bU(\bP)$ and $\bF(\bP)$ requires raising and
lowering of indices followed by direct substitution in equation
(\ref{TMUNUMHD}) to find the components of the MHD stress-energy
tensor.

Since we update $\bU$ rather than $\bP$, we must solve for $\bP(\bU)$ at
the end of each timestep.  We use a multidimensional Newton-Raphson
routine with the value of $\bP$ from the last timestep as an initial
guess.  Since $B^i$ can be obtained analytically, only 5 equations need
to be solved.  The Newton-Raphson method requires an expensive
evaluation of the Jacobian $\partial \bU/\partial \bP$.  In practice we
evaluate the Jacobian analytically.  It is possible to evaluate the
Jacobian by numerical derivatives, but this is both expensive and a
source of numerical noise.  

The evaluation of $\bP(\bU)$ is at the heart of our numerical scheme;
the procedure must be robust.  We have found that it is crucial that the
errors (differences between the current and target values of $\bU$) used
to evaluate convergence in the Newton-Raphson scheme be properly
normalized.  We normalized the errors with $\detg \rho u^t$.

To evaluate $\bF$ we use a MUSCL type scheme with ``HLL'' fluxes
\citep{hll83}.  The fluxes are defined at zone faces.  A
slope-limited linear extrapolation from the zone center gives
$\bP_R$ and $\bP_L$, the primitive variables at the right and
left side of each zone interface.  We have implemented the
monotonized central (``Woodward'', or ``MC'') limiter, the van
Leer limiter, and the minmod limiter; unless otherwise stated,
the tests described here use the MC limiter, which is the least
diffusive of the three.

From $\bP_R, \bP_L$, calculate the maximum left and rightgoing wave
speeds $c_{\pm,R}, c_{\pm,L}$, and the fluxes $\bF_R = \bF(\bP_R)$ and
$\bF_L = \bF(\bP_L)$.  Defining $c_{max} \equiv {\rm MAX}(0, c_{+,R},
c_{+,L})$ and $c_{min} \equiv -{\rm MIN}(0, c_{-,R},c_{-,L})$, the HLL
flux is then
\begin{equation}\label{HLLFLUX}
\bF = {c_{min} \bF_R + c_{max} \bF_L - c_{max} c_{min}
(\bU_R - \bU_L)\over{c_{max} + c_{min}}}
\end{equation}
If $c_{max} = c_{min}$, the HLL flux becomes the so-called local
Lax-Friedrichs flux.  

\subsection{Constrained Transport}

The pure HLL scheme will not preserve any numerical representation of
$\bnabla \cdot \bB = 0$.  An incomplete list of options for handling
this constraint numerically includes: (1) ignore the production of
monopoles by truncation error and hope for the best (in our experience
this causes the scheme to fail in any complex flow); (2) introduce a
divergence-cleaning step (this entails solving an elliptic equation at
each timestep); (3) use an \cite{eh88} type constrained transport scheme
(this requires a staggered mesh, so that the magnetic field components
are zone face centered); (4) introduce a diffusion term that causes
numerically generated monopoles to diffuse away (\cite{mar87}; this
typically leaves a monopole field with rms value somewhat larger than
the truncation error).  

We have chosen a fifth option, a version of constrained transport that
can be used with a zone-centered scheme.  This idea was introduced by
one of us in \cite{toth00}, where it is called the flux-interpolated
constrained transport (or ``flux-CT'') scheme \footnote{We have also
experimented with T\'oth's ``flux-CD'', which preserves a different
representation of $\bnabla \cdot \bB = 0$.  This appears to be slightly
less robust.  It also has a larger effective stencil.}.  It preserves a
numerical representation of $\bnabla \cdot \bB = 0$ by smoothing the
fluxes with a special operator.  The disadvantage of this method is that
it is more diffusive than the ``bare'', unconstrained scheme.  The
advantage is that it is extremely simple.

To clarify how zone-centered constrained transport works we now give a
specific example for a special relativistic problem in Cartesian
coordinates $t,x,y$.  To fix notation, write the induction equation as
\begin{equation}
\del_t B^i = -\del_j F^j_i,
\end{equation}
where we have used $\detg = 1$ and the fluxes $F^j_i$ are
\begin{equation}
\begin{array}{rcl}
F^x_x & = & 0 \\
F^y_y & = & 0 \\
F^y_x & = & b^y u^x - b^x u^y \\
F^x_y & = & b^x u^y - b^y u^x = -F^y_x 
\end{array}
\end{equation}
Notice that the fluxes are centered at different locations on the grid:
$F^x$ fluxes live on the $x$ face of each zone at grid location
$i-1/2,j$ (we use $i,j$ to denote the center of each zone), while $F^y$
fluxes live on the $y$ face at $i,j-1/2$.  The smoothing operator
replaces the numerical (HLL-derived) $F_i^j$ with
$\tilde{F}^j_i$, defined by 
\begin{equation}
\begin{array}{rcl}
\tilde{F}^x_x(i-1/2,j) & = & 0 \\
\tilde{F}^y_y(i,j-1/2) & = & 0 \\
\tilde{F}^y_x(i, j-1/2) & = & {1\over{8}} \Bigl[ 2 F^y_x(i, j-1/2) \\
 & & + F^y_x(i+1, j-1/2) + F^y_x(i-1, j-1/2) \\
 & & - F^x_y(i-1/2, j) - F^x_y(i+1/2, j) \\ 
 & & - F^x_y(i-1/2, j-1) - F^x_y(i+1/2, j-1) \Bigr] \\
\tilde{F}^x_y(i-1/2,j) & = & {1\over{8}} \Bigl[ 2 F^x_y(i-1/2, j) \\
 & & + F^x_y(i-1/2, j+1) + F^x_y(i-1/2, j-1) \\
 & & - F^y_x(i, j-1/2) - F^y_x(i, j+1/2) \\
 & & - F^y_x(i-1, j-1/2) - F^y_x(i-1, j+1/2) \Bigr]
\end{array}
\end{equation}
It is a straightforward but tedious exercise to verify that this
preserves the following corner-centered numerical representation of
$\bnabla \cdot \bB$:
\begin{equation}
\begin{array}{rcl}
\bnabla \cdot \bB 
& = &   \left( B^x(i,j) + B^x(i,j-1) - B^x(i-1,j) -
B^x(i-1,j-1)\right)/(2 \Delta x) \\
&   & + \left(B^y(i,j) + B^y(i-1,j) - B^y(i,j-1) - 
B^y(i-1,j-1)\right)/(2 \Delta y),
\end{array}
\end{equation}
where $\Delta x$ and $\Delta y$ are the grid spacing.  

\subsection{Wave Speeds}

The HLL approximate Riemann solver does not require eigenvectors of the
characteristic matrix (as would a Roe-type scheme), but it does require
the maximum and minimum wave speed (eigenvalues).  These wave speeds are
also required to fix the timestep via the Courant conditions.  The
relevant speed is the phase speed ``$\omega/k$'' of the wave, and it
turns out that only speeds for waves with wavevectors aligned along
coordinate axes are required.  Suppose, for example, that one needs to
know how rapidly signals propagate in the fluid along the $x_1$
direction.  First, find a wavevector $k_\mu = (-\omega, k_1,0,0)$, that
satisfies the dispersion relation for the mode in question: $D(k_\mu) =
0$.  Then the wave speed is simply $\omega/k_1$.

The dispersion relation $D(k_\mu) = 0$ for MHD waves has a simple form
in a comoving frame.  In terms of the relativistic sound speed $c_s^2 =
(\del (\rho + u)/\del p)_s^{-1} = \gamma p/w$, (the last holds only if
$p = (\gamma - 1) u$) and the relativistic Alfv\'en velocity $\bva =
\bB/\sqrt{\sE}$, where $\sE = b^2 + w$ and $w \equiv \rho + u + p$, the
dispersion relation is
\begin{equation}
\begin{array}{l}
\omega \left(\omega^2 - \kdv^2\right) \times \\
\left(\omega^4 + \omega^2 \, 
	\left( k^2 (\va^2 + c_s^2 (1 - \va^2/c^2)) 
	+ c_s^2 \kdv^2/c^2 \right)
	+ k^2 c_s^2 \kdv^2\right) = 0,
\end{array}
\end{equation}
Here $c$ is the (temporarily reintroduced) speed of light.  The first
term is the zero frequency entropy mode, the second is the Alfv\'en
mode, and the third contains the fast and slow modes.  The eighth mode
is eliminated by the no-monopoles constraint.

The relativistic sound speed asymptotes to $c \sqrt{\gamma - 1} =
c/\sqrt{3}$ for $\gamma = 4/3$, and the Alfv\'en speed asymptotes to
$c$.  In the limit that $B^2/\rho \gg 1$ and $p/\rho \sim \ll 1$, the
GRMHD equations are a superset of the time-dependent, force-free
electrodynamics equations recently discussed by \cite{kom02b}; these
contain fast modes and Alfv\'en modes that move with the speed of light.
They are indistinguishable from vacuum electromagnetic modes only when
their wavevector is oriented along the magnetic field.

To find the maximum wave speeds we need to evaluate the comoving-frame
dispersion relation for the fast wave branch from coordinate frame
quantities.  This is straightforward because the dispersion relation
depends on scalars, which can be evaluated in any frame: $\omega = k_\mu
u^\mu$; $k^2 = K_\mu K^\mu$, where $K_\mu = (g_{\mu\nu} + u_\mu u_\nu)
k^\nu$ is the part of the wavevector normal to the fluid 4-velocity;
$\va^2 = b_\mu b^\mu/\sE$; $\kdv = k_\mu b^\mu/\sqrt{\sE}$.  The
relevant portion of the dispersion relation (for fast and slow modes) is
thus a fourth order polynomial in the components of $k_\mu$.  This can
be solved either analytically or by standard numerical methods.  The two
fast mode speeds are then used as $c_{max}$ and $c_{min}$ in the HLL
fluxes.

We have found it convenient to replace the full dispersion relation by
an approximation:
\begin{equation}
\omega^2 = (\va^2 + c_s^2 (1 - \va^2/c^2)) k^2.
\end{equation}
This overestimates the maximum wavespeed by a factor $\le 2$ in the
comoving frame.  The maximum error occurs for $\bk \parallel \bva,
v_A/c_s = 1,$ and $v_A \ll c$, and it is usually much less, particularly
if the fluid is moving super-Alfv\'enically with respect to the grid.
This approximation is convenient because it is quadratic in $k_\mu$, and
so can be solved more easily.

\subsection{Implementation Notes}

For completeness we now give some details of the implementation of the
algorithm.

{\bf Time Stepping.}  Our scheme is made second order in time by taking
a half-step from $t^n$ to $t^{n+1/2}$, evaluating $\bF(\bP(t^{n+1/2}))$,
and using that to update $\bU(t^n)$ to $\bU(t^{n+1})$.  

{\bf Modification of Energy Equation.}  A direct implementation of the
energy equation can be inaccurate because the magnetic and internal
energy density can be orders of magnitude smaller than the rest mass
density.  To avoid this we subtract the particle number conservation
equation from the energy equation, i.e., we evolve
\begin{equation}
\del_t (\detg (T^t_t + \rho u^t)) = -\del_i (\detg (T^i_t + \rho u^i))
	+ \detg T^\kappa_\lambda \Gamma^\lambda_{t\kappa}.
\end{equation}
In the nonrelativistic limit, this procedure subtracts the rest mass
energy density from the total energy density.

{\bf Specification of Geometric Quantities.}  In two dimensions we need
to evaluate $g_{\mu\nu}, g^{\mu\nu},$ and $\detg$ at four points in
every grid zone (the center, two faces, and one corner) and
$\Gamma^\mu_{\nu\lambda}$ at the zone center.  It would be difficult to
accurately encode analytic expressions for all these quantities.  HARM
is coded so that an analytic expression need only be provided for
$g_{\mu\nu}$; all other geometric quantities are calculated numerically.
The connection, for example, is obtained to sufficient accuracy by
numerical differentiation of the metric.  This minimizes the risk of
coding errors in specifying the geometry.  It also minimizes coordinate
dependent code, making it relatively easy to change coordinate systems.
Minimal coordinate dependence, besides following the spirit of general
relativity, enables one to perform a sort of fixed mesh refinement by
adapting the coordinates to the problem at hand.  For example, near a
Kerr black hole we use $\log(r)$ as the radial coordinate instead of the
usual Boyer-Lindquist $r$, and this concentrates numerical resolution
toward the horizon, where it is needed.

{\bf Density and Internal Energy Floors.}  Negative densities and
internal energies are forbidden by the GRMHD equations, but numerically
nothing prevents their appearance.  In fact, negative internal energies
are common in numerical integrations with large density or pressure
contrast.  Following common practice, we prevent this by introducing
``floor'' values for the density and internal energy.  These floors are
enforced after the half-step and the full step.  They preserve velocity
but do not conserve rest mass or energy-momentum.

{\bf Outflow Boundary Conditions.}  In the rotating black hole
calculations described below we use outflow boundary conditions at the
inner and outer radial boundaries.  The usual implementation of outflow
boundary conditions is to simply copy the primitive variables from the
boundary zones into the ghost zones.  This can result in unphysical
values of the primitive variables in the ghost zones-- for example,
velocities that lie outside the light cone-- because of variations in
the metric between the boundary and ghost zones.  

We have experimented with a variety of schemes for projecting variables
into the ghost zones in the context of black hole accretion flow
calculations (described in \S 5).  We find that some are more robust
than others.  The most robust extrapolates the density, internal energy,
and radial magnetic field according to
\begin{equation}
P({\rm ghost}) = P({\rm boundary}) \detg({\rm boundary})/\detg({\rm ghost}),
\end{equation}
the $\theta$ and $\phi$ components of the velocity and magnetic field
according to 
\begin{equation}
P({\rm ghost}) = P({\rm boundary}) (1. - \Delta r/r),
\end{equation}
and the radial velocity according to
\begin{equation}
P({\rm ghost}) = P({\rm boundary}) (1. + \Delta r/r).
\end{equation}
The extrapolation of $\theta$ and $\phi$ components of magnetic field
and velocity results in weak damping of these components near the
boundary.  Slightly different choices of the extrapolation coefficients
(i.e. $(1. - 2 \Delta r/r)$) are much less robust.

{\bf Performance.}  We have implemented both serial and parallel
versions of the code.  In serial mode the code integrates the black hole
accretion problem (described in \S 5) at $\approx 54,000$ zone cycles
per second on a 2.4 GHz Intel Pentium 4, when compiled using the Intel C
compiler.  The parallel code was implemented using MPI.  

\section{Code Verification}

Here we present a test suite for verifying a GRMHD code.  The tests are
nonrelativistic, special and general relativistic, and one and two
dimensional.  The list of problems for which there are known, exact
solution is short, since exact solutions of multidimensional GRMHD
problems are algebraically complicated.  This list of test problems was
developed in collaboration with J. Hawley and J.-P. de Villiers.  Unless
otherwise stated we set $\gamma = 4/3$ and $c = 1$.

\subsection{Linear Modes}

This first test considers the evolution of a small amplitude wave in two
dimensions.  The unperturbed state is $\rho = 1$, $p = 1$, $u^i = 0$,
$B^y = B^z = 0$, $B^x = B^x_0$.  The basic state is parametrized by
$\alpha = (B^x_0)^2/(\rho c^2)$; our fiducial test runs have $\alpha =
1$.  Onto this basic state we introduce a perturbation of the form
$\exp(i \bk \cdot {\bf x} - i \omega(\bk) t)$, where $(k_x,k_y) = (2\pi,
2\pi)$, and the amplitude is fixed by $\delta B^y = 10^{-4} B^x_0$.  The
computational domain is $x,y \in [0,1),[0,1)$, and the boundary
conditions are periodic.  The wave is either slow, Alfv\'enic, or fast.  

This test exercises almost all terms in the governing equations.  The
numerical resolution is $(N_x,N_y) \equiv N (5,4)$ zones, and the
integration runs for a single wave period $2\pi/\omega$, so that a
perfect scheme would return the simulation to its original state.  We
measure the $\Lone$ norm of the difference between the final state and
the initial state for each primitive variable.  For example, we measure
\begin{equation}
\Lone(\delta\rho) = \int dx dy |\rho({\rm t = 0}) - 
	\rho({\rm t = 2\pi/\omega})|.
\end{equation}
for the density.

\begin{figure}
\plotone{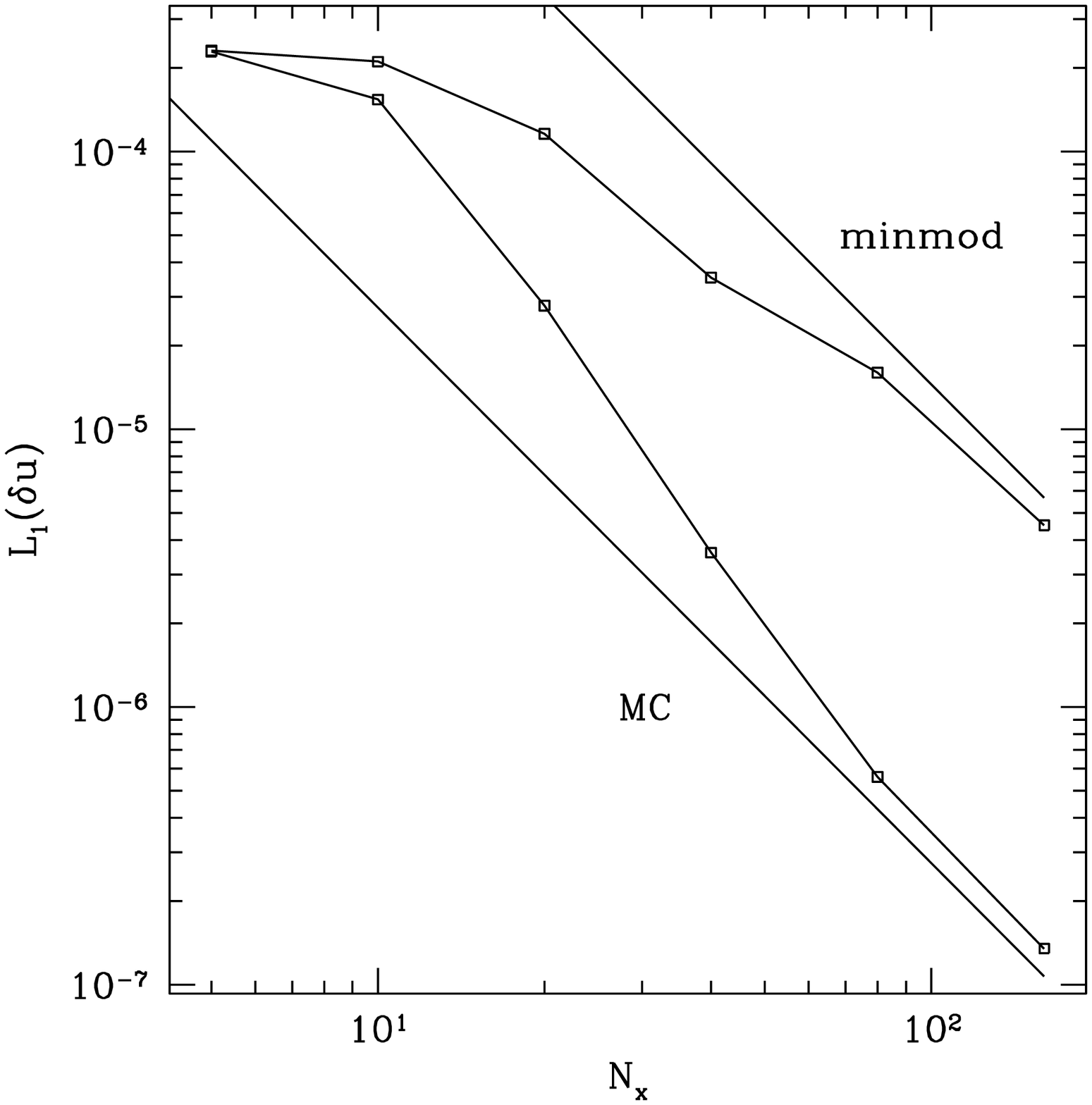}
\caption{
The $\Lone$ norm of the error in $u$ for a slow
wave as a function of $N_x$ for both the monotonized central (MC) and
minmod limiter.  The straight lines show the slope expected for 
second order convergence.
}\label{fig-B}
\end{figure}

\begin{figure}
\plotone{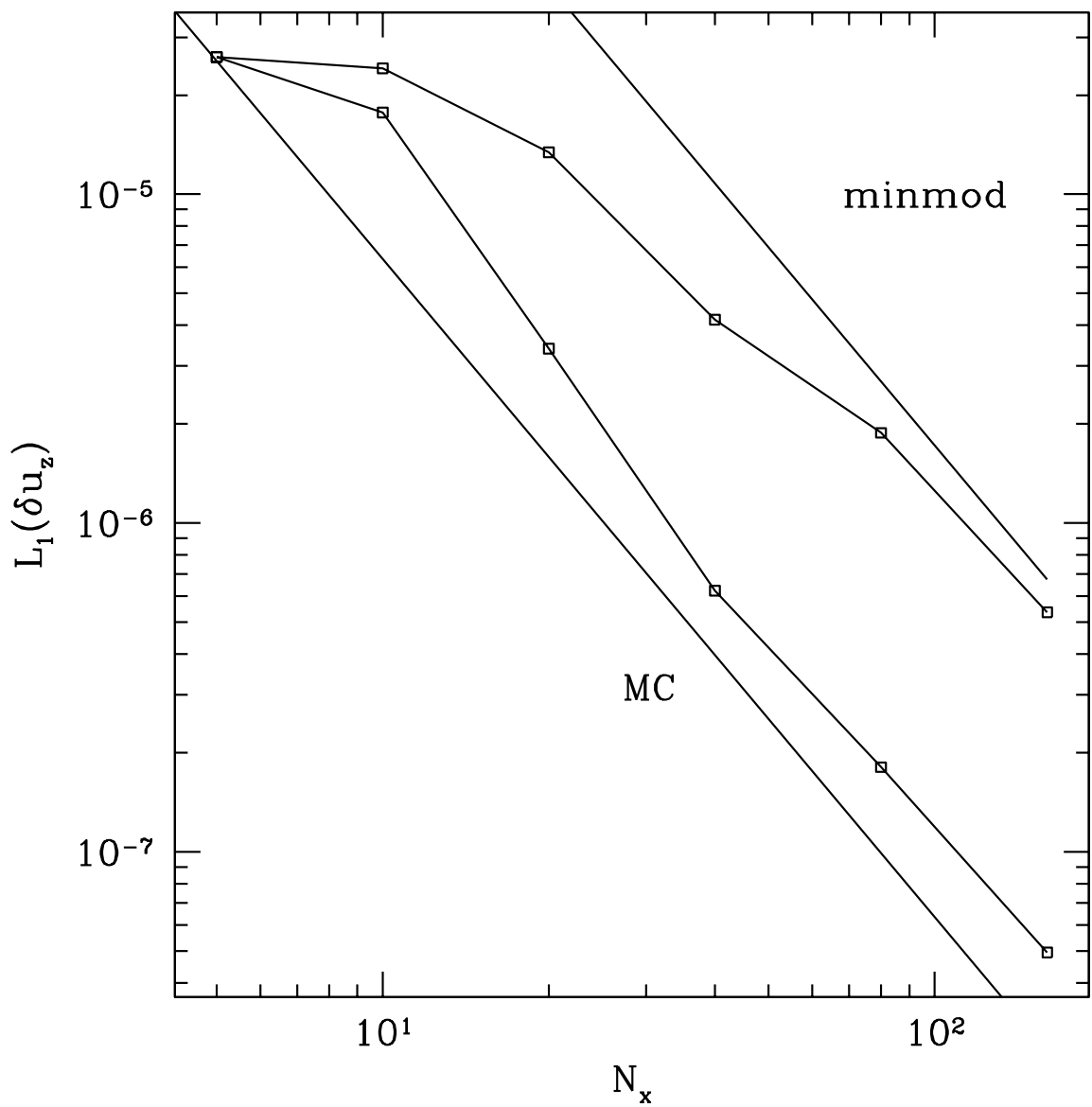}
\caption{
The $\Lone$ norm of the error in the single nonzero
component of the velocity for an \alf wave as a function of $N_x$ for
both the monotonized central (MC) and minmod limiter.  The straight
lines show the slope expected for second order convergence.
}\label{fig-C}
\end{figure}

\begin{figure}
\plotone{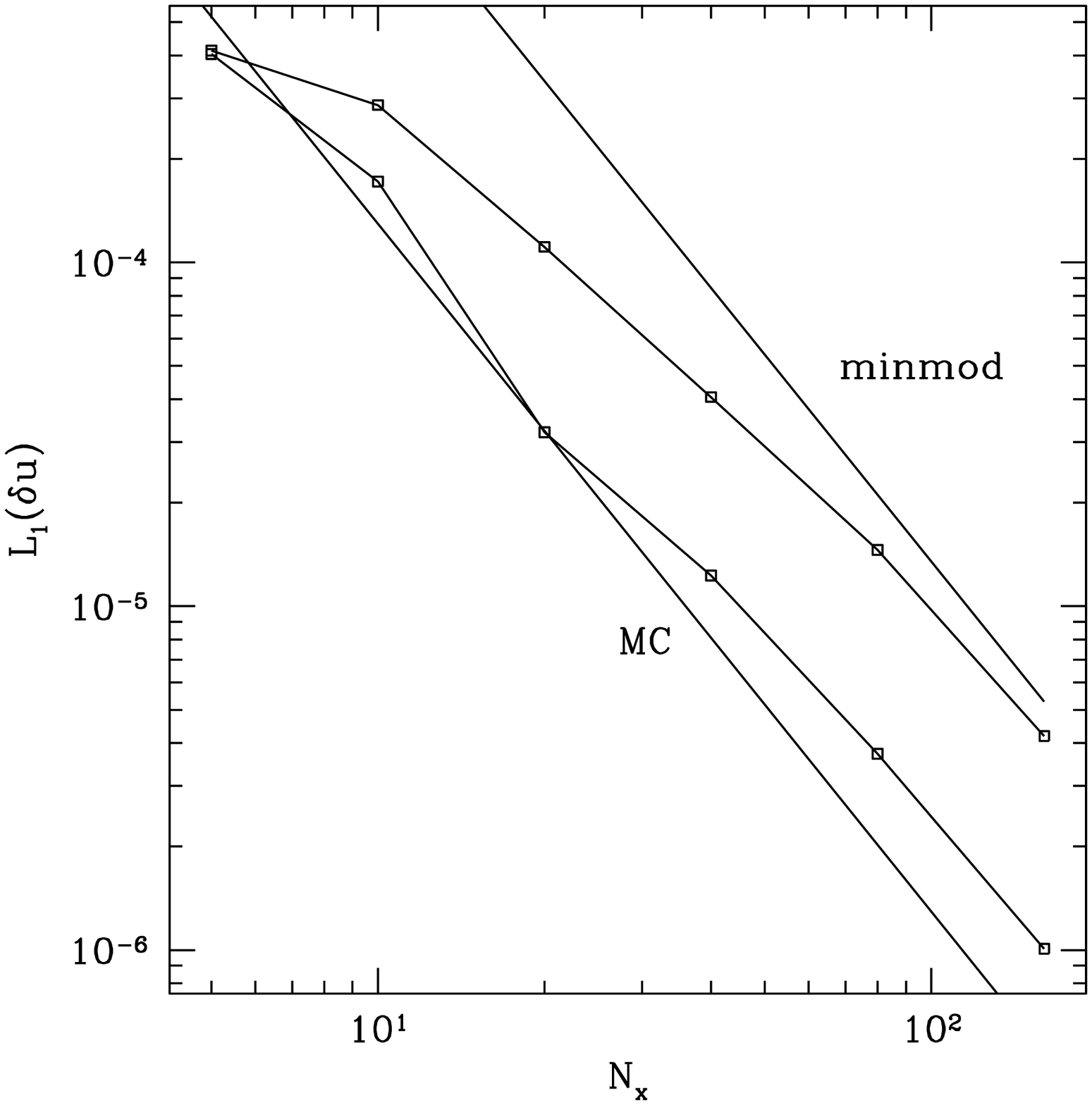}
\caption{
The $\Lone$ norm of the error in $u$ for a fast
wave as a function of $N_x$ for both the monotonized central (MC) and
minmod limiter.  The straight lines show the slope expected for 
second order convergence.
}\label{fig-D}
\end{figure}

All primitive variables exhibit similar convergence properties
(as they must, since with the exception of the magnetic field,
they are tightly coupled together).  In Figures \ref{fig-B},
\ref{fig-C}, and \ref{fig-D} we present the $\Lone$ norm of the
error for runs using the monotonized central limiter and a
Courant number of $0.8$, in addition to the results for the
minmod limiter.  These runs have $\alpha = 1$.  Figure
\ref{fig-B} shows the results for the slow wave, Figure
\ref{fig-C} for the \alf wave, and Figure \ref{fig-D} for the
fast wave.  Evidently the convergence rate asymptotes to second
order, although more slowly for the minmod limiter.

The code performs similarly well over a range of $\alpha$, provided only
that $\delta B^2/2 \ll p$, which is necessary for the wave to be in the
linear regime.  We have been unable to find a value of $B^x_0$ where the
code fails completely for a linear amplitude disturbance, although for
very large values of $B^x_0$ the evolution becomes inaccurate because of
numerical noise in the evaluation of $\bP(\bU)$.

\subsection{Nonlinear Waves}

\cite{kom99} has proposed a suite of one-dimensional nonlinear
tests for special relativistic MHD.   Komissarov presents a total
of 9 tests (see his Table 1).  The nonlinear Alfv\'en wave (test
5), and the compound wave (test 6) cannot be reconstructed
without a separate derivation of the exact analytic solution, and
we will not provide that here.  For the remaining tests
Komissarov's Table 1 contains several misprints that are
corrected in \cite{kom02a}.  Our code is able to integrate each
of Komissarov's remaining 7 tests, although in some cases we must
reduce the Courant number (usually $0.8$) or resort to the
slightly more robust van Leer slope limiter.  Tests that required
special treatment are: fast shock (Courant number $= 0.5$); shock
tube 1 (Courant number $=0.3$, van Leer limiter); shock tube 2
(Courant number $= 0.5$); collision (Courant number $= 0.3$; van
Leer limiter).  

\begin{figure}
\plotone{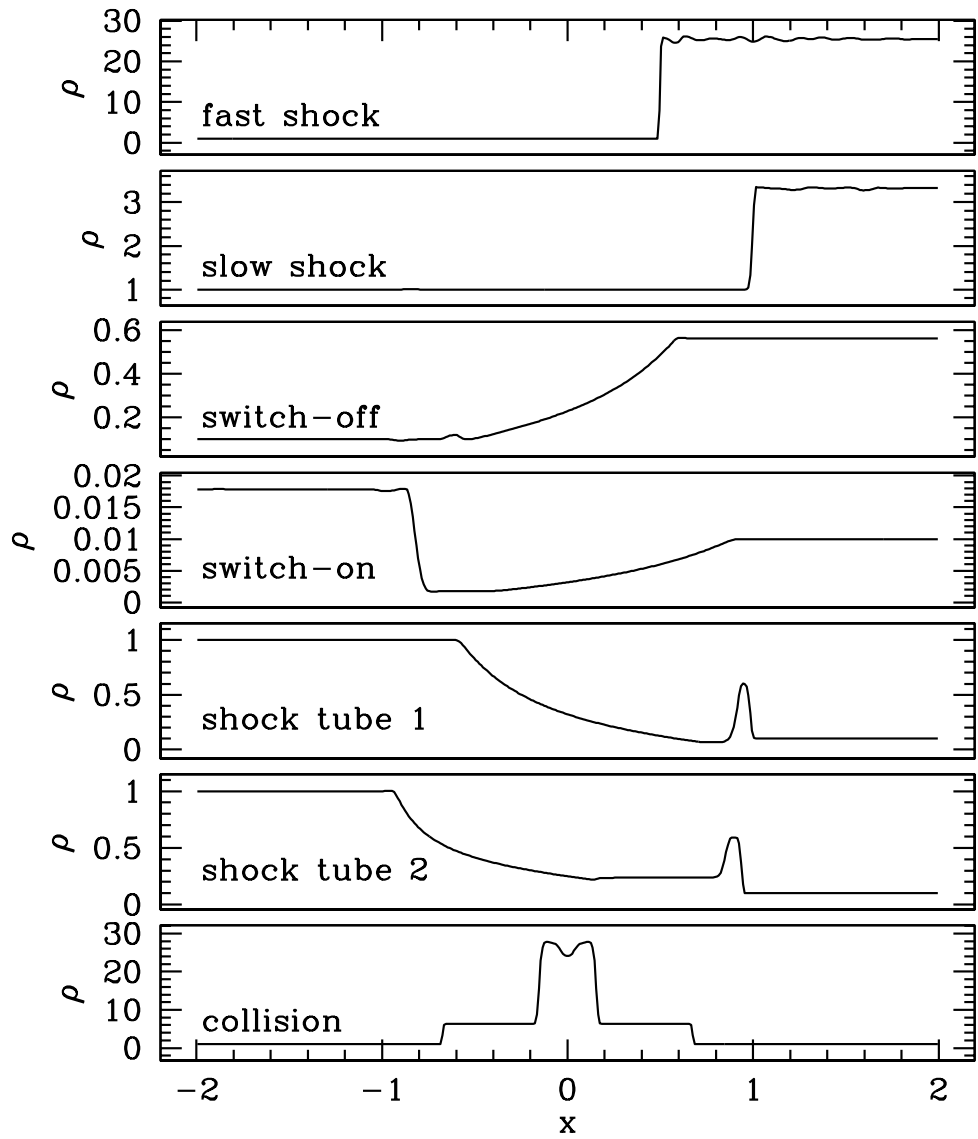}
\caption{
The run of density in the Komissarov nonlinear wave tests.
}\label{fig-E}
\end{figure}

\begin{figure}
\plotone{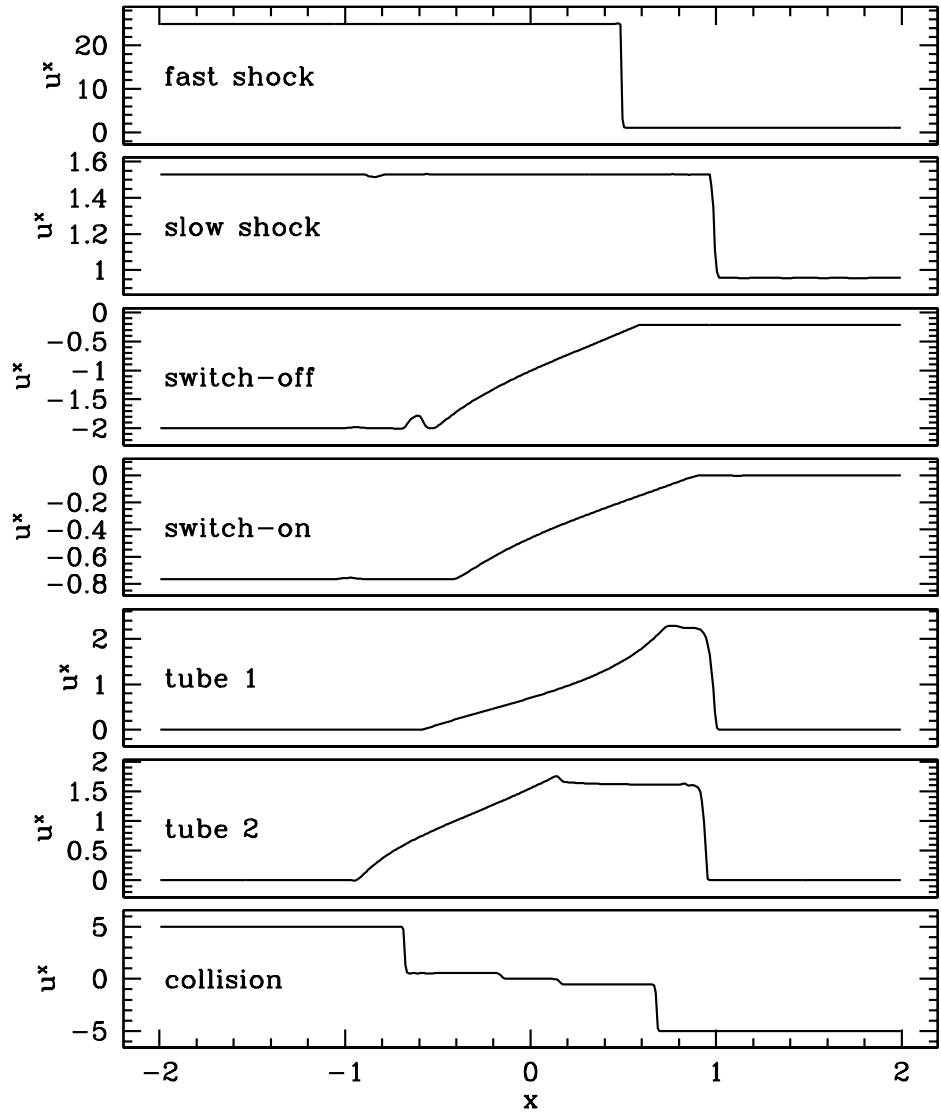}
\caption{
The run of $u^x$ in the Komissarov nonlinear wave tests.
}\label{fig-F}
\end{figure}

Figures \ref{fig-E} and \ref{fig-F} show the run of $\rho$ and $u^x$,
respectively, for all 7 tests.  These may be compared with Komissarov's
figures.  Notice that, unlike Komissarov, we have in all cases set $N_x
= 400$ and $x \in (-2,2)$.  We have not obtained the exact solutions
used by Komissarov, but the solutions can still be checked
quantitatively.  For example, the slow shock wave speed is $0.5$; since
the calculation ends at $t = 2$ the slow shock front should be, and is,
located at $x \approx 1.0$.  The fast shock speed is $0.2$, so at $t =
2.5$ the fast shock wave front should be, and is, located at $x \approx
0.5$.

There are artifacts evident in the figures.  In particular there is
ringing near the base of the switch-on and switch-off rarefaction waves.
This is common and is seen in Komissarov's results as well.  In addition
the narrow, Lorentz-contracted shell of material behind the shock in
shock tube 1 is poorly resolved; the correct shell density is $\approx
0.88$ but a resolution of $N_x > 800$ is required to find this result to
an accuracy of a few percent.  There is also a transient associated with
the fast shock that propagates off the grid and so is not visible in
Figures \ref{fig-E} and \ref{fig-F}.

A complete set of nonlinear wave tests for one dimensional {\it
nonrelativistic} MHD was developed by \cite{rj95} (hereafter RJ).  We
can run these under HARM by rescaling the speed of light to $c = 10^2$
in code units, where all velocities in the tests are $\order(1)$.  This
should lead to results that agree with RJ to $\order(v/c) \approx 1\%$.
The results can be checked quantitatively by comparison to the tables
provided by RJ.

\begin{figure}
\plotone{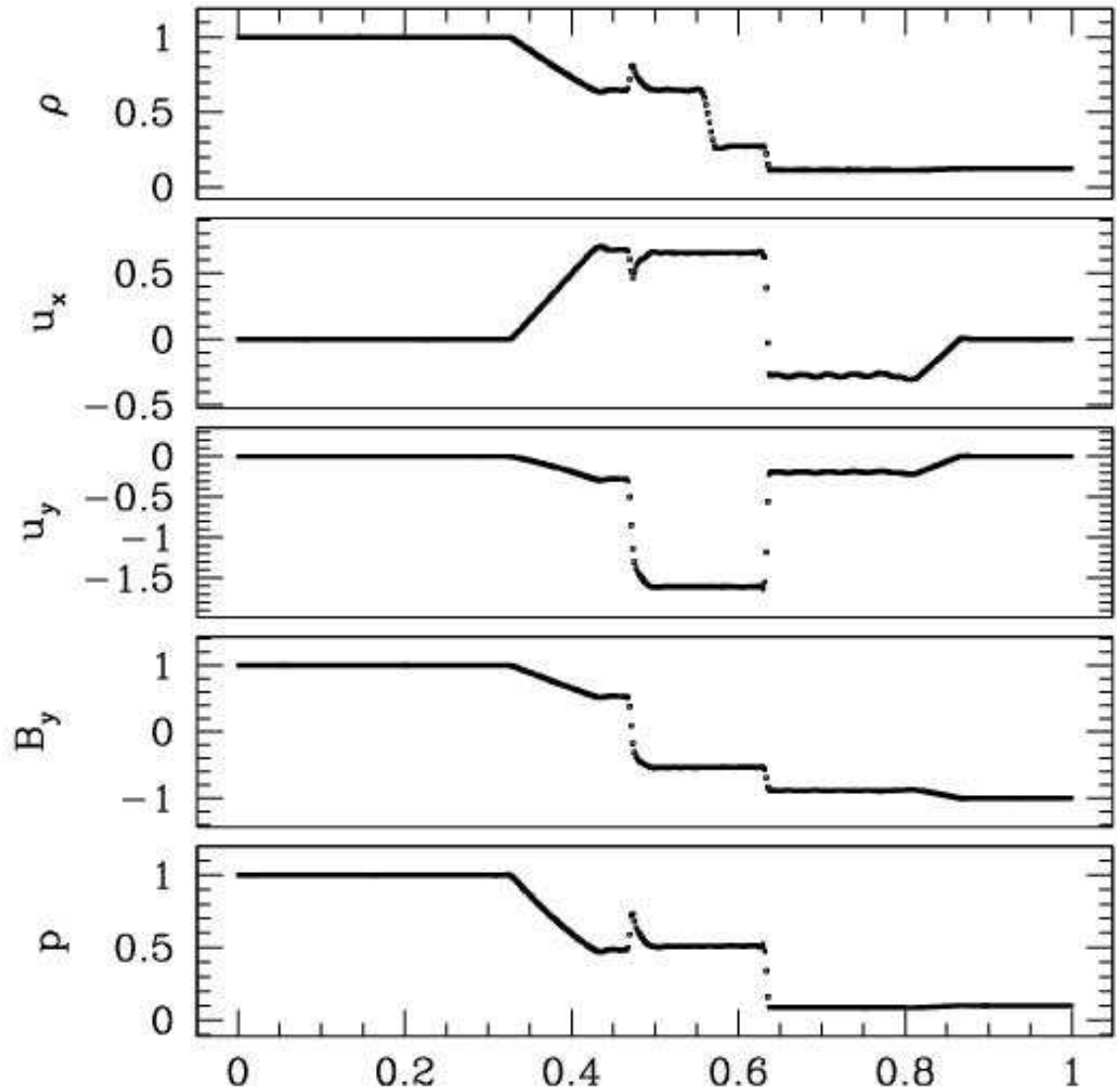}
\caption{
Snapshot of the final state in HARM's integration of Ryu \& Jones test
5A (a version of the Brio \& Wu shock tube) but with $c = 100$.  The
figure shows primitive variable values at $t = 0.15$.  Quantitative
agreement is found to within $\approx 1\%$, as expected.
}\label{fig-G}
\end{figure}

Figure \ref{fig-G} shows our results for RJ test 5A, which is a version
of the familiar \cite{bw88} magnetized shock tube test.  Like RJ we use
$512$ zones between $x = 0$ and $x = 1$, and we measure the results at
$t = 0.15$.  To compare to RJ quantitatively, consider $u_x$ in the
region behind the fast rarefaction wave, near $x = 0.7$.  RJ report $u_x
= -0.277$ here, while we measure $u_x = -0.273$, which differs by $1\%$,
as expected.  Similar agreement is found for the other variables.  The
most unsatisfactory feature of the solution is the visible post-shock
oscillations.  The amplitude of these features varies, depending on the
Courant number (here $0.9$) and the choice of slope limiter (here MC).

\begin{figure}
\plotone{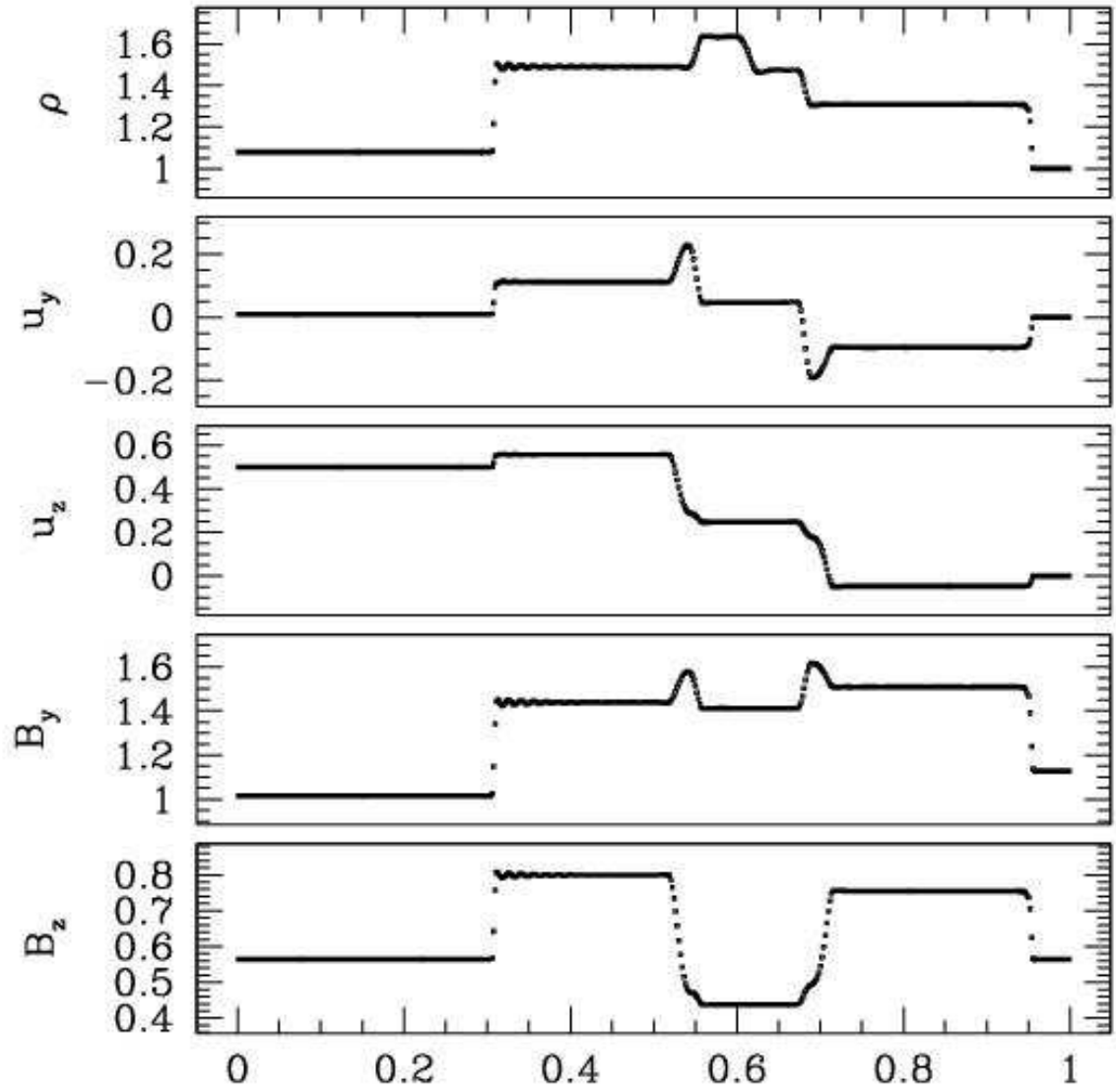}
\caption{
Snapshot of the final state in HARM's integration of Ryu \& Jones test 2A,
with $c = 100$.
}\label{fig-H}
\end{figure}

Figure \ref{fig-H} shows our results for RJ test 2A.  In the
region near $x = 0.6$, RJ report that $B_y = 1.4126$, and we find
essentially exact agreement ($B_y = 1.41262$) after averaging
over a small region near $x = 0.6$.  This test has two pairs of
closely spaced slow shocks and rotational discontinuities that
are difficult to resolve, and our scheme barely obtains the
correct peak values of $u_y$ and $B_y$, even though there are
about $20$ zones inside the ``horns'' visible in the $B_y$ panel
of the figure.

\subsection{Transport}

This special relativistic test evolves a disk of enhanced density moving
at an angle to the grid until it returns to its original position.  The
computation is carried out in a domain $x,y \in [-0.5,0.5),[-0.5,0.5)$
and the boundary conditions are periodic.  The initial state has $v^x =
v^y = 0.7$, or $u^x = u^y \approx 4.95$, corresponding to $u^t \approx
7.07$.  The initial density $\rho = 1$ except in a disk at $r < r_{s} =
0.45$, where $\rho = 3/2 + \cos(2 \pi r/r_s)$.  The initial pressure $p
= 1$, and the initial magnetic field is zero.   The test is run until $t
= 10/7$, when the system should return exactly to its initial state.

\begin{figure}
\plotone{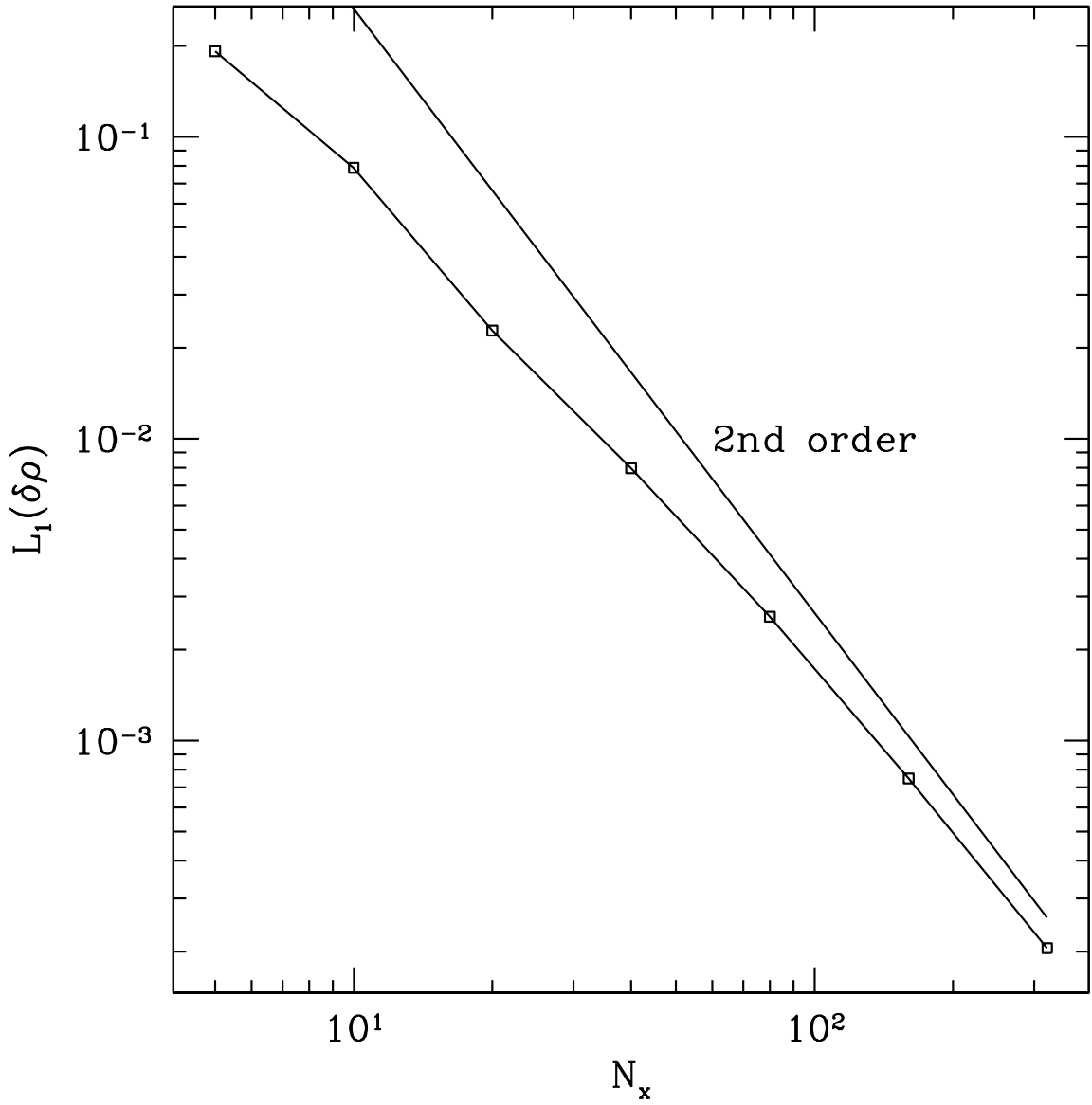}
\caption{
Convergence results for the transport test.  
}\label{fig-A}
\end{figure}

Numerically, we use the monotonized central limiter and set the Courant
number to $0.8$.  The resolution is fixed so that $N_x = 5 N_y/4$.
Figure \ref{fig-A} shows the $L_1$ norm of the error in $\rho$ as a
function of $x$ resolution.  The convergence rate asymptotes to second
order.

\subsection{Orszag-Tang Vortex}

The Orszag-Tang vortex (OTV) is a classic nonlinear MHD problem
\citep{ot79}.  Here we compare our code, with the speed of light set to
$100$ so that that it is effectively nonrelativistic, to the output of
VAC \citep{toth96}, an independent nonrelativistic code developed by one
of us.  The version of VAC used here is TVD-MUSCL using the monotonized
central limiter.  It is dimensionally unsplit and uses a scheme similar
to HARM to control $\bnabla\cdot\bB$.  The problem is integrated in the
periodic domain $x \in (-\pi,\pi]$, $y \in (-\pi,\pi]$ from $t = 0$ to
$t = \pi$.  Our version of the OTV has $\gamma = 4/3$, but is otherwise
identical to the standard problem.

\begin{figure}
\plotone{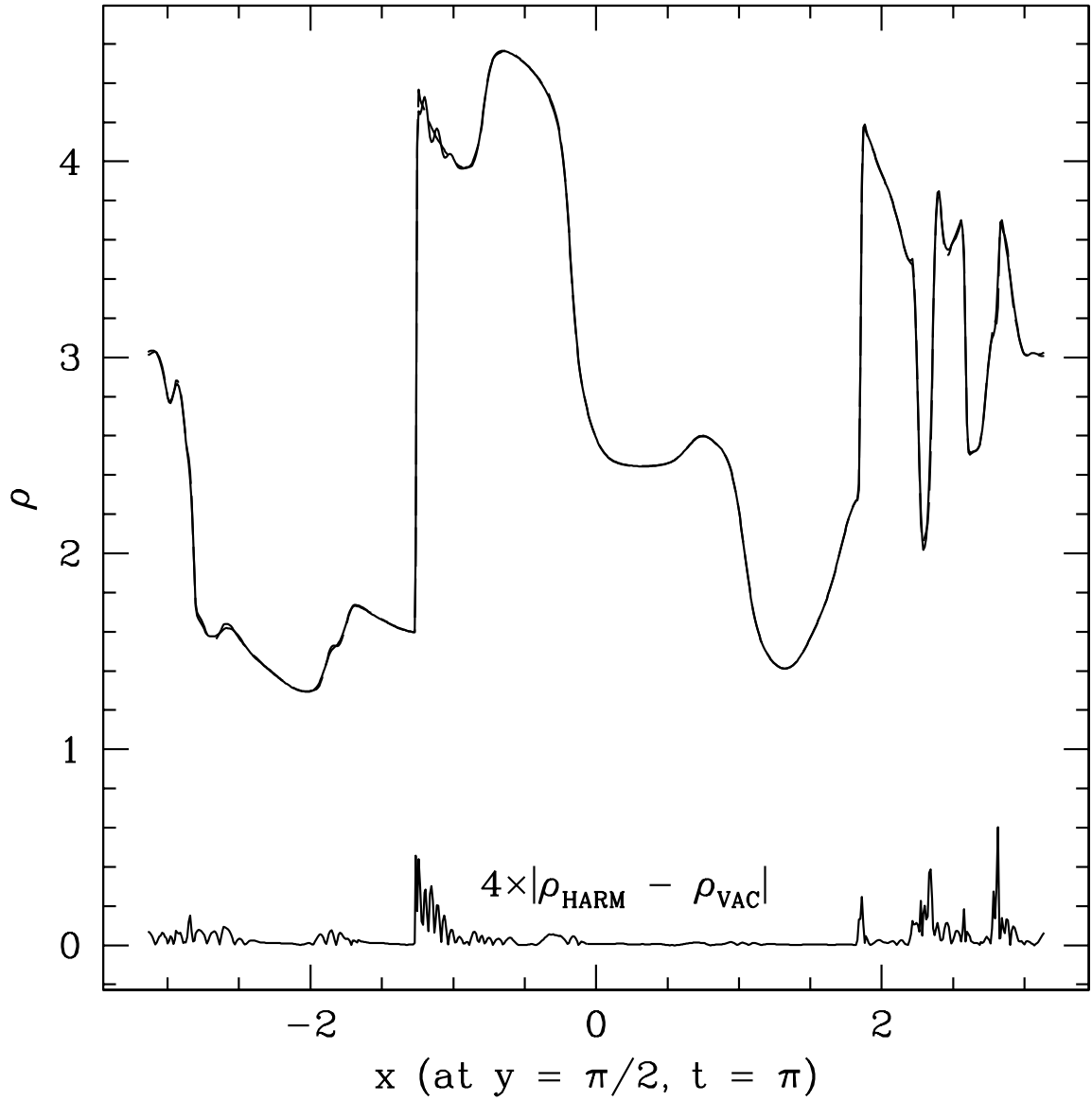}
\caption{
A cut through the density in the nonrelativistic Orszag-Tang vortex
solution from HARM (solid line, with $c = 100$), from VAC (dashed line),
and $4 \times $ the difference (lower solid line) at a resolution of
$640^2$.
}\label{fig-N}
\end{figure}

Results are shown in Figure \ref{fig-N}, which shows $\rho$ along a cut
through the model at $y = \pi/2$ and $t = \pi$.  The resolution is
$640^2$.  The solid line shows the results from HARM; the dashed line
shows the results from VAC.  The lower solid line shows the difference
between the two multiplied by $4$.  Evidently our code behaves similarly
to VAC on this problem.  

\begin{figure}
\plotone{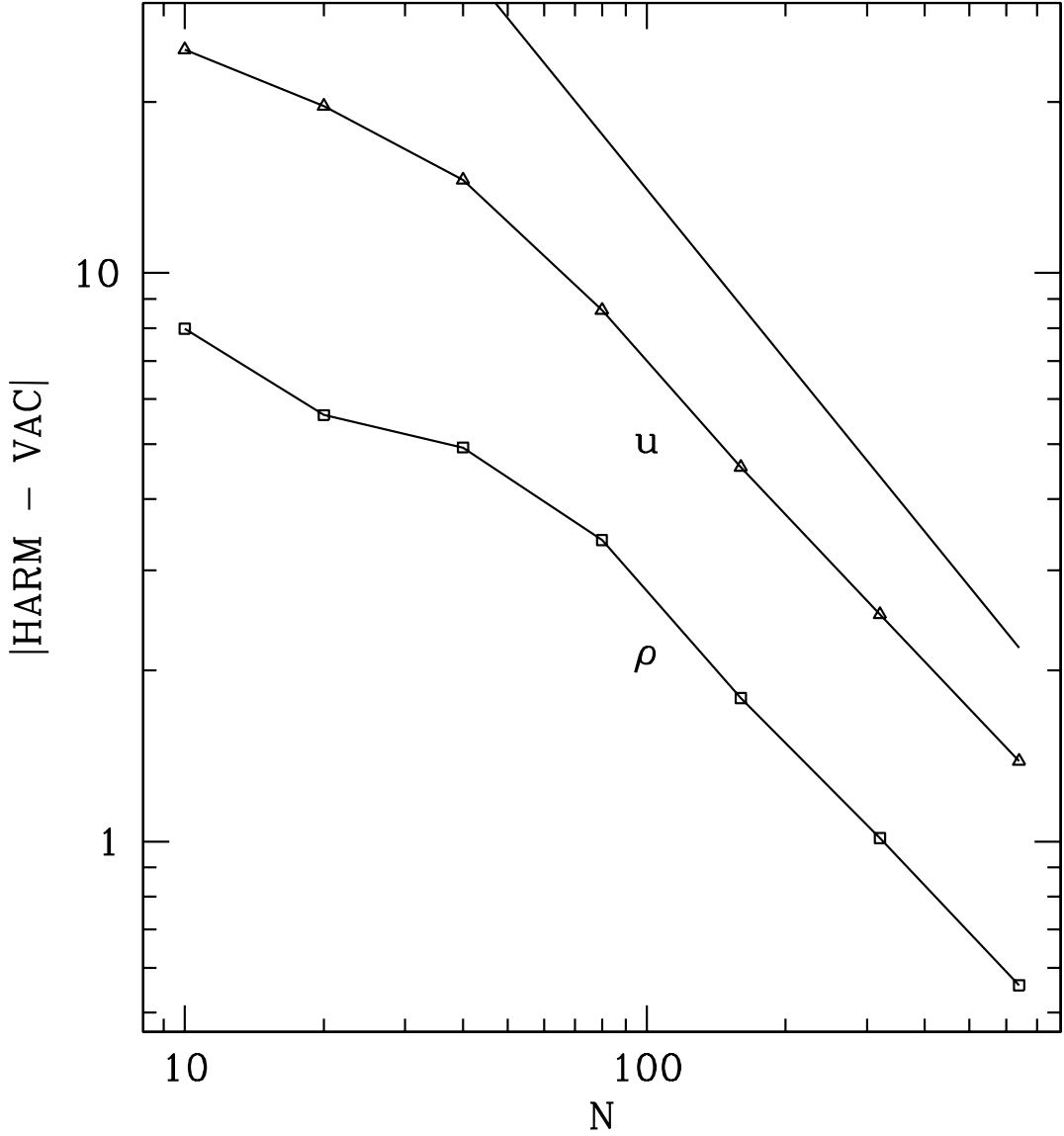}
\caption{
Comparison of results from HARM and the nonrelativistic MHD code VAC for
the Orszag-Tang vortex.  The plot shows the $\Lone$ norm of the
difference between the two results as a function of resolution for the
primitive variables $\rho$ (squares) and $u$ (triangles).  The straight
line shows the slope expected for first order convergence.  The errors
are large because they are an integral over an area of $(2\pi)^2$.
}\label{fig-P}
\end{figure}

We can quantify this by asking how the difference between the HARM and
VAC solutions changes as a function of resolution.  Figure \ref{fig-P}
shows the variation in the $\Lone$ norm of the difference between the
two solutions.  Thus the line marked $\rho$ shows
\begin{equation}
\int dx dy |\rho({\rm HARM; N^2}) - \rho({\rm VAC; N^2})|
\end{equation}
evaluated at $t = \pi$.  The codes converge to one another approximately
linearly, as expected for a flow containing discontinuities.  If this
study were extended to higher resolution convergence would eventually
cease because the HARM solution would differ from the VAC solution due
to finite relativistic corrections.

\subsection{Bondi Flow in Schwarzschild Geometry}

Spherically symmetric accretion (Bondi flow) in the Schwarzschild
geometry has an analytic solution (see, e.g., \cite{st83}) that can be
compared with the output of our code.  This appears to be a
one-dimensional test, but for HARM it is actually two dimensional.
Although the pressure is independent of the Boyer-Lindquist coordinate
$\theta$, the $\theta$ acceleration does not vanish identically.  This
is because pressure enters the momentum equations through a flux
($-\del_\theta (p \sin\theta)$ in the Newtonian limit) and a source term
($p \cos\theta$ in the Newtonian limit).  Analytically these terms
cancel; numerically they produce an acceleration that is of order the
truncation error.  

Our test problem follows that set out in \cite{hsw}: we fix the sonic
point $r_s = 8 G M/c^2$, $\dot{M} = 4\pi r^2 \rho u^r =  -1$, and
$\gamma = 4/3$.  The problem is integrated in the domain $r \in (1.9,20)
G M/c^2$ for $\Delta t = 100 G M/c^3$.  We use coordinates based on the
Kerr-Schild system, whose line element is
\begin{equation}\label{KSLINE}
\begin{array}{ccl}
ds^2 & = & -(1 - 2 r/\rho^2) dt^2 + (4 r/\rho^2) dr dt + (1 + 2 r/\rho^2) dr^2
	+ \rho^2 d\theta^2 + \\
	& & \sin^2\theta \left(\rho^2 + a^2 (1 + 2
	r/\rho^2) \sin^2\theta\right) d\phi^2 \\
	& & - (4 a r\sin^2\theta/\rho^2 ) dtd\phi - 2 a (1 + 2 r/\rho^2) 
	\sin^2\theta dr d\phi,
\end{array}
\end{equation}
where we have set $G M = c = 1$.  In (\ref{KSLINE}) only $\rho^2
= r^2 + a^2 \cos^2(\theta)$; elsewhere $\rho$ is density.  In
this test, $a = 0$.  We modify these coordinates by replacing $r$
by $x_1 = \log(r)$.  The new coordinates are implemented by
changing the metric rather than changing the spacing of grid
zones.  We measure the $\Lone$ norm of the difference between the
initial conditions (exact analytic solution) and the final state.
The difference is taken over the inner $3/4$ of the grid in each
direction, thus excluding boundary zones where errors may scale
differently.  This test exercises many terms in the code because
in Kerr-Schild coordinates only three of the ten independent
components of the metric are zero.

\begin{figure}
\plotone{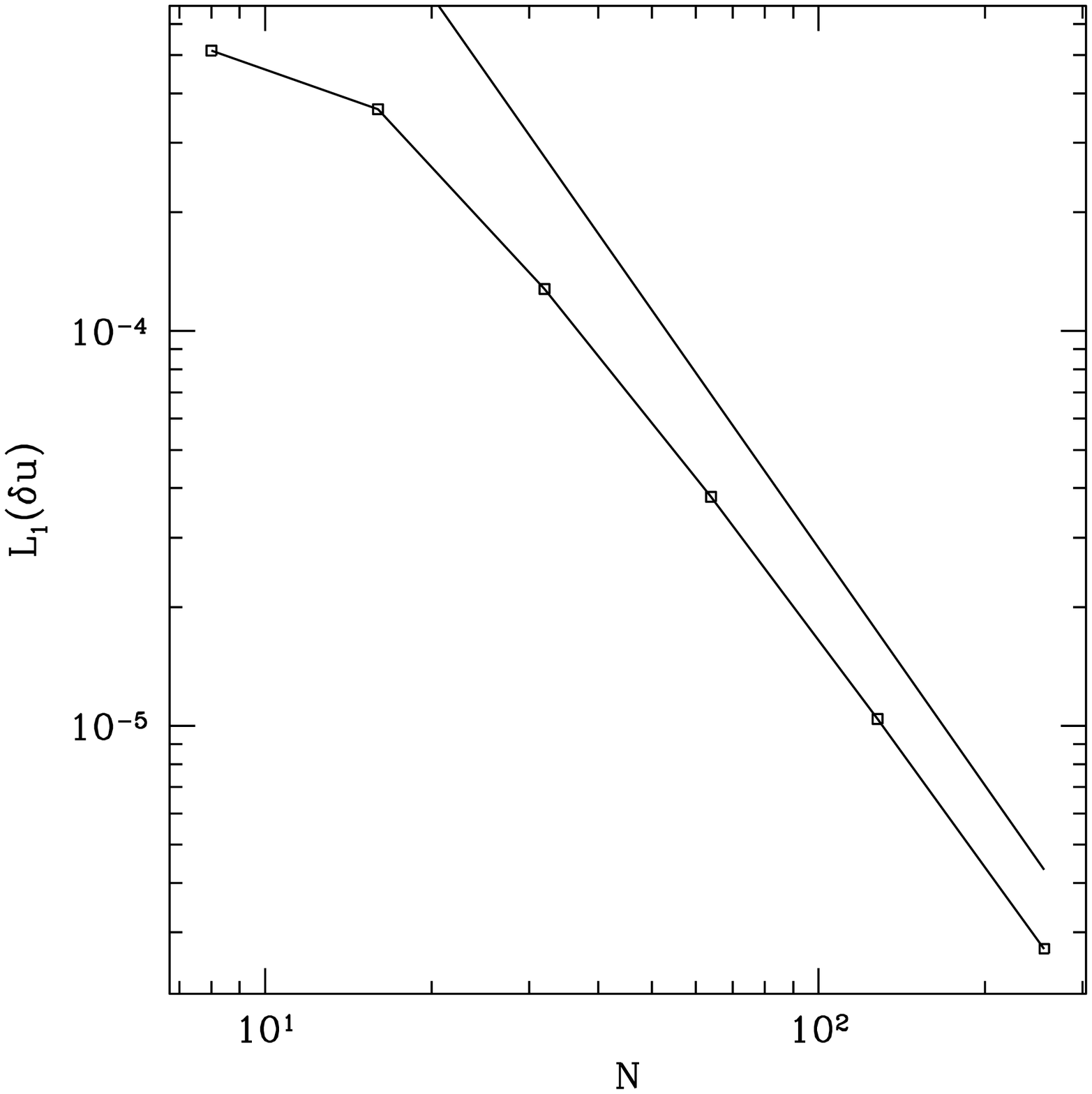}
\caption{
Convergence results for the unmagnetized Bondi accretion test onto
a Schwarzschild black hole.  The straight line shows the slope 
expected for second order convergence.
}\label{fig-I}
\end{figure}

The $\Lone$ norm of the error in internal energy for the Bondi test is
shown in Figure \ref{fig-I}.  Similar results obtain for the other
independent variables.  The solution converges at second order.

\subsection{Magnetized Bondi Flow}\label{MBF}

The next test considers a Bondi flow containing a spherically symmetric,
radial magnetic field.  The solution to this problem is identical to the
Bondi flow described above because the flow is along the magnetic field,
so all magnetic forces cancel exactly.  This is a difficult test,
however, because numerically the magnetic terms cancel only to
truncation error.  This causes problems at high magnetic field strength.  

\begin{figure}
\plotone{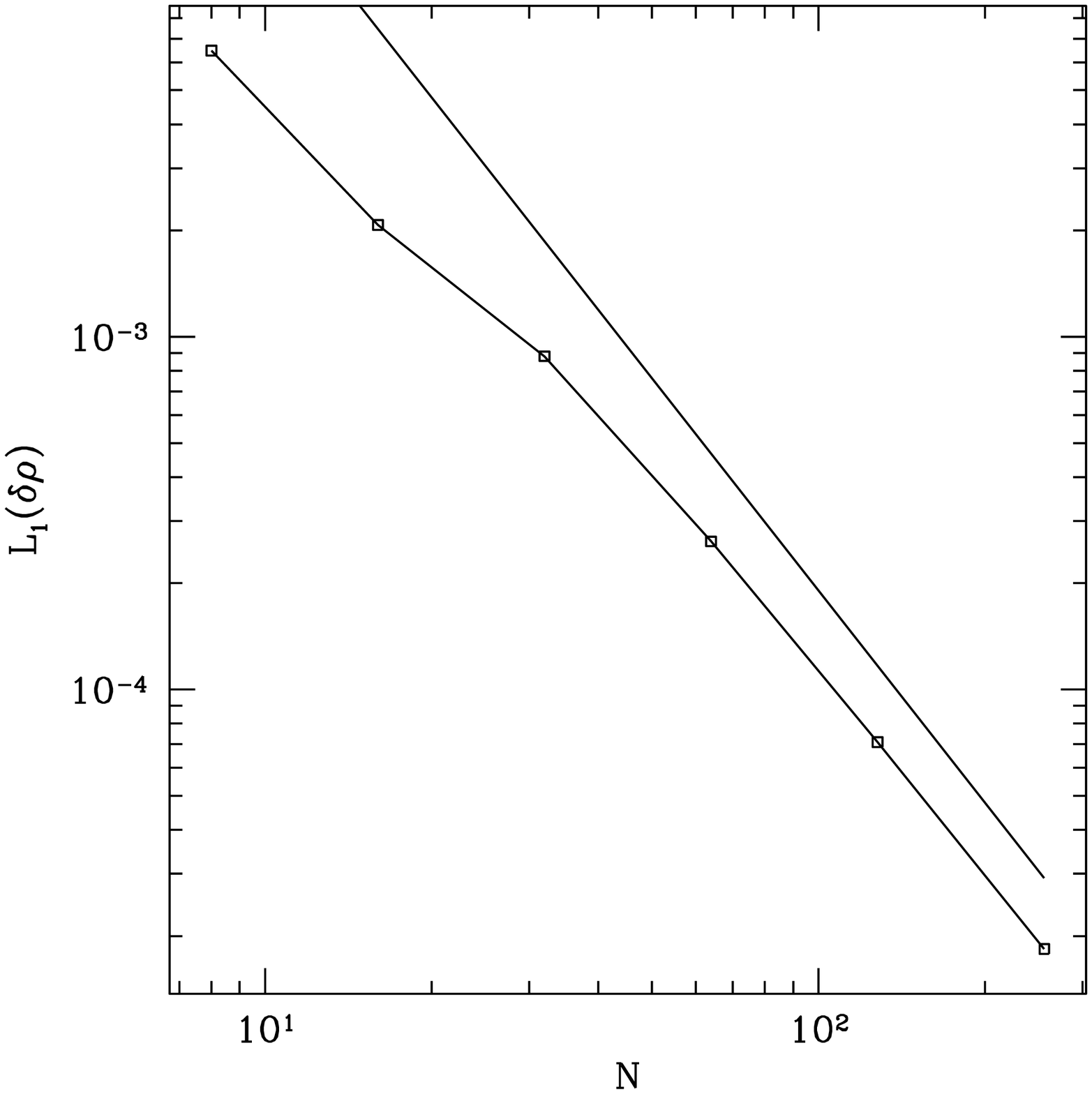}
\caption{
Convergence results for the magnetized Bondi accretion test onto
a Schwarzschild black hole.  The straight line shows the slope 
expected for second order convergence.
}\label{fig-J}
\end{figure}

We use the same Bondi solution as in the last subsection and
parameterize the magnetic field strength by $b^2/\rho$ at the inner
boundary.  Our fiducial run has $(b^2/\rho)(r = 1.9 G M/c^2) = 10.56$.
The $\Lone$ norm of the error in the internal energy is shown in Figure
\ref{fig-J}.  Similar results obtain for the other independent
variables.  The solution converges at second order.

We have considered models with a range of $(b^2/\rho)(r_{in})$.
Lowering $(b^2/\rho)(r_{in})$ produces results similar to those in our
fiducial test run.  Raising $(b^2/\rho)(r_{in})$ first causes the code
to produce inaccurate results (at $\sim 10^3$, where the radial velocity
profile is smoothly distorted from the true solution) and then to fail
(at $\sim 10^4$).  This is an example of a general problem with
conservative schemes when the basic energy density scales (rest mass,
magnetic, and internal) differ by many orders of magnitude.   

\subsection{Magnetized Equatorial Inflow in Kerr Geometry}

This test considers the steady-state, magnetized inflow solutions found
by \cite{tak90}, as specialized to the case of inflow inside the
marginally stable orbit by \cite{gam99}.  This solution exercises many
of the important terms in the governing equations, in particular the
interaction of the magnetized fluid with the Kerr geometry.  

We use Boyer-Lindquist coordinates to specify this problem, but the
solution is integrated in the modified Kerr-Schild coordinates described
above.  The flow is assumed to lie in the neighborhood of the black
hole's equatorial plane and is thus one dimensional, much like the
Weber-Davis model for the solar wind.  As above, we set $G M = c = 1$.

\begin{figure}
\plotone{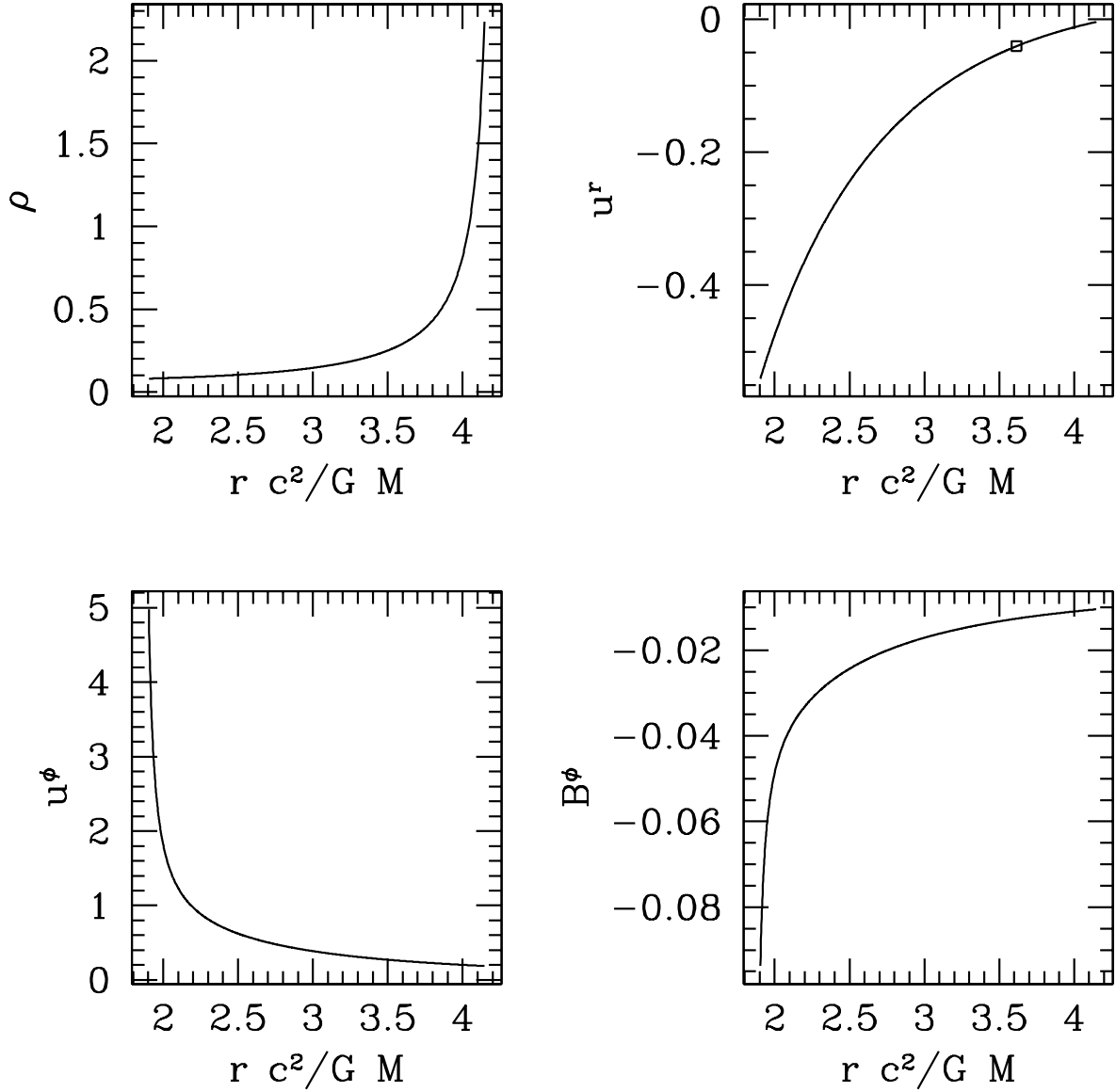}
\caption{
The equatorial inflow solution in the Kerr metric for $a/M = 0.5$
and magnetization parameter $F_{\theta\phi} = 0.5$.  The panels show
density, radial component of the four-velocity in Boyer-Lindquist
coordinates (with the square showing the location of the fast point),
the $\phi$ component of the four-velocity, and the toroidal magnetic
field $B^\phi = F^{\phi t}$.
}\label{fig-Q}
\end{figure}

The particular inflow solution we consider is for a black hole
with spin parameter $a/M = 0.5$.  The model has an accretion rate
$F_M = -1 = 2\pi\rho r^2 u^r$ (adopting the notation of Gammie
1999).  The magnetization parameter $F_{\theta\phi} = r^2 B^r =
0.5$.  The flow is constrained to match to a circular orbit at
the marginally stable orbit.  This is enough to uniquely specify
the flow.  It follows that (see Gammie 1999) $F_{t\theta} =
\Omega F_{\theta\phi}$, where $\Omega$ is the orbital frequency
at the marginally stable orbit.  For $a/M = 0.5, \Omega \approx
0.10859$.  The solution that is regular at the fast point has
angular momentum flux $F_L = 2\pi r^2 (u^r u_\phi
- b^r b_\phi) \approx -2.8153$ and energy flux $F_E = 2\pi r^2
  (u^r u_t
- b^r b_t) \approx -0.90838$.  The fast point is located at $r
  \approx 3.6167$, and the radial component of the four-velocity
  there is $u^r = -0.040547$.  Figure \ref{fig-Q} shows the
  radial run of the solution.

We initialize the flow with a numerical solution that is subject to
roundoff error.  The near-equatorial nature of the solution is mimicked
by using a single zone in the $\theta$ direction centered on $\theta =
\pi/2$.  The computational domain runs from $1.02 \times$ the horizon
radius $r_h$ to $0.98 \times$ the radius of the marginally stable orbit
$r_{mso}$.  For $a/M = 0.5$, $r_h = 1.866$, and $r_{mso} = 4.233$.  The
analytic flow model is cold (zero temperature) but we set the initial
internal energy in the code equal to a small value instead.  The model
is run for $\Delta t = 15$.

\begin{figure}
\plotone{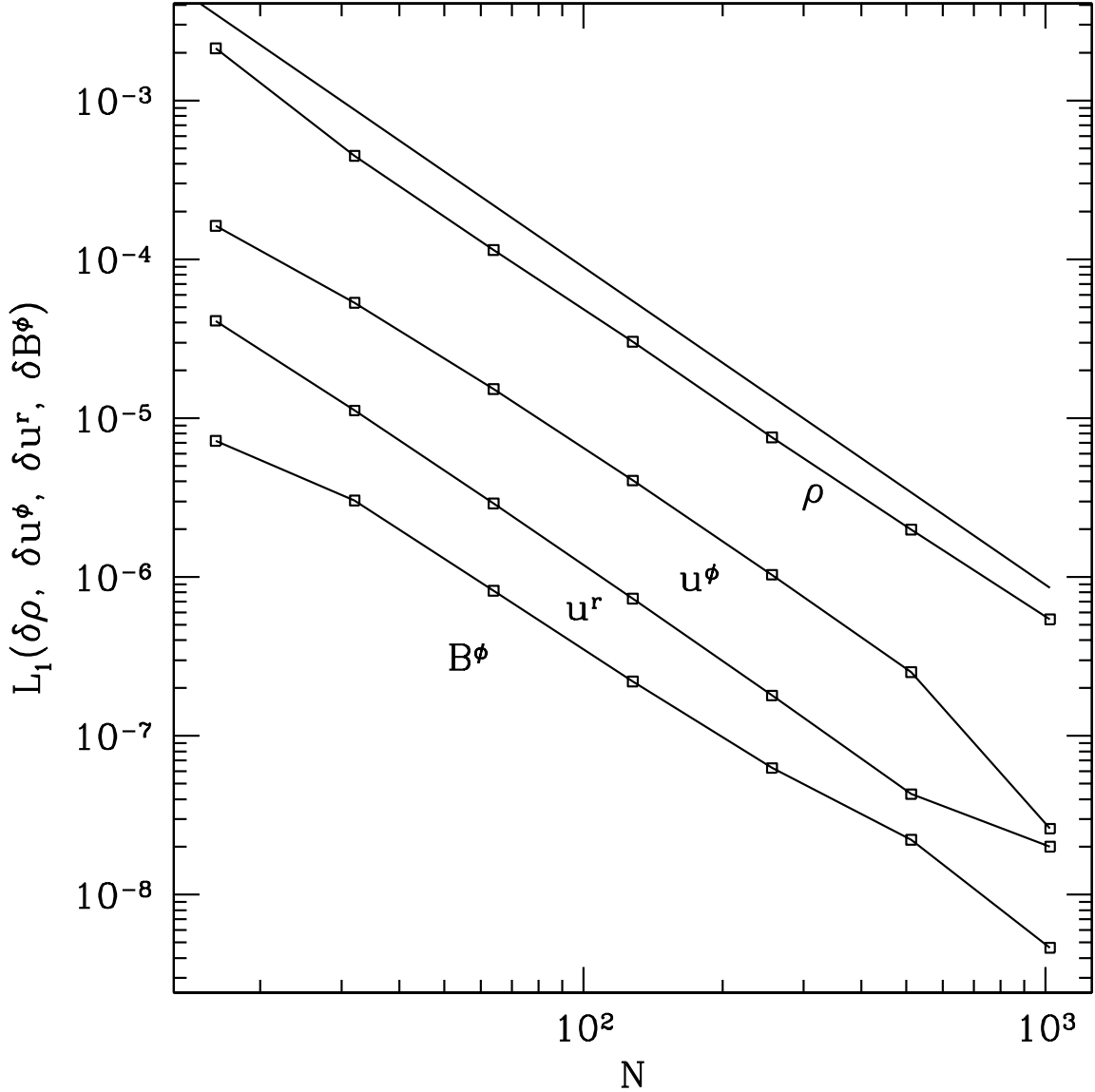}
\caption{
Convergence results for the magnetized inflow solution in a Kerr metric
with $a/M = 0.5$.  Parameters for the initial, quasi-analytic solution
are given in the text.  The straight line shows the slope expected for
second order convergence.  The $\Lone$ error norm for each of the
nontrivial variables are shown.  The small deviation from second order
convergence at high resolution is due to numerical errors in the
quasi-analytic solution used to initialize the solution.
}\label{fig-K}
\end{figure}

Figure \ref{fig-K} shows the $\Lone$ norm of the error in $\rho, u^r,
u^\phi,$ and $B^\phi$ and a function of the total number of radial
gridpoints $N$.  The straight line shows the slope expected for second
order convergence.  The small deviation from second order convergence at
high $N$ in several of the variables is due to numerical errors in the
initial solution, which relies on numerical derivatives \citep{gam99}.

\subsection{Equilibrium Torus}

Our next test concerns an equilibrium torus.  This class of equilibria,
found originally by \cite{fm76} and \cite{ajs78}, consist of a ``donut''
of plasma surrounding a black hole.  The donut is supported by both
centrifugal forces and pressure and is embedded in a vacuum.  Here we
consider a particular instance of the Fishbone \& Moncrief solution.

A practical problem with this test is that HARM abhors a vacuum.  We
have therefore introduced floors on the density and internal energy that
limit how small these quantities can be.  The floors are dependent on
radius, with $\rho_{min} = 10^{-4} (r/r_{in})^{-3/2}$ and $u_{min} =
10^{-6} (r/r_{in})^{-5/2}$.  This means that the torus is surrounded by
an insubstantial, but dynamic, accreting atmosphere that interacts with
the torus surface.  To minimize the influence of the atmosphere on our
convergence test, we take the $\Lone$ norm of the change in variables
only over that region where $\rho > 0.02 \rho_{max}$.

The problem is integrated in modified Kerr-Schild coordinates.  The
Kerr-Schild radius $r$ has been replaced by the logarithmic radial
coordinate $x_1 = \ln(r)$, and the Kerr-Schild latitude $\theta$ has
been replaced by $x_2$ such that $\theta = \pi x_2 + (1/2)(1 -
h)\sin(2\pi x_2)$.  Clearly $0 \le x_2 \le 1$ maps to $0 \le \theta \le
\pi$.  This coordinate transformation has a single adjustable parameter
$h$; for $h = 1$ we recover the original coordinate system (the $\theta$
coordinate is simply rescaled by $\pi$).  As $h \rightarrow 0$ numerical
resolution is concentrated near the midplane.  

\begin{figure}
\plotone{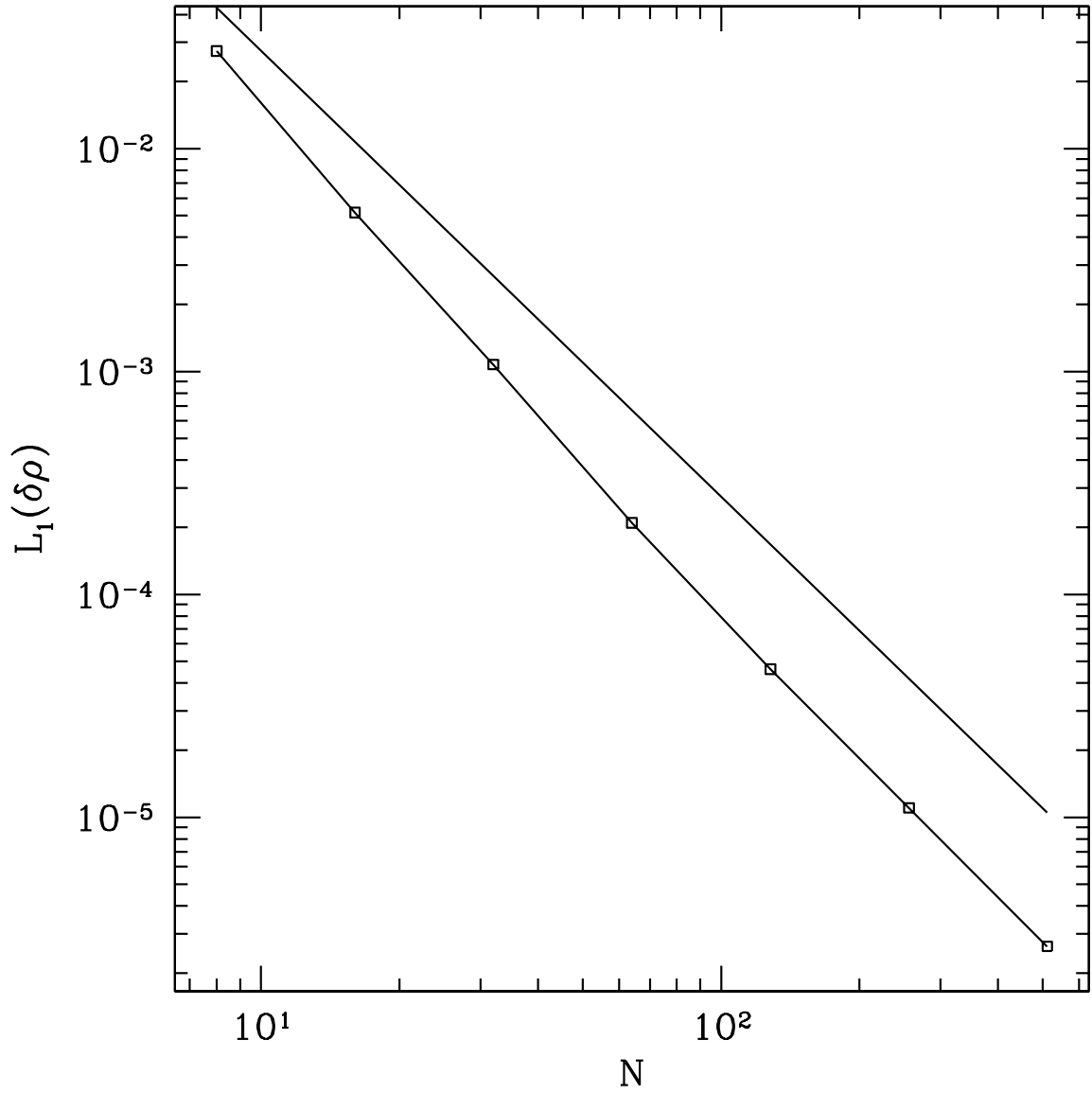}
\caption{
Convergence results for the Fishbone and Moncrief equilibrium disk
around an $a/M = 0.95$ black hole. 
}\label{fig-L}
\end{figure}

We have integrated a Fishbone-Moncrief disk around a black hole with
$a/M = 0.95$, to maximize general relativistic effects.  We set $u^t
u_\phi = const. = 3.85$ (this is the defining feature of the
Fishbone-Moncrief equilibria) and $r_{in} = 3.7$.  The grid extends
radially from $R_{in} = 0.98 r_h = $ to $R_{out} = 20$.  The coordinate
parameter $h$ described in the last paragraph is set to $0.2$.  The
numerical resolution is $N \times N$, where $N = 8, 16, 32, \ldots,
512$, and the solution is integrated for $\Delta t = 10$.  Figure
\ref{fig-L} shows the $\Lone$ norm of the error for each variable as a
function of $N$.  Second order convergence is obtained.  

The sum of the evidence presented in this section strongly suggests that
we are solving the equations of GRMHD without significant, compromising
errors.

\section{Magnetized Torus Near Rotating Black Hole}

Finally we offer an example of how HARM can be applied to a real
astrophysical problem: the evolution of a magnetized torus near a
rotating black hole.  Again we set $G M = c = 1$.

The initial conditions contain a Fishbone-Moncrief torus with $a/M =
0.5$, $r(p_{max}) = 12$, and $r_{in} = 6$.  Superposed on this
equilibrium is a purely poloidal magnetic field with vector potential
$A_\phi \propto {\rm MAX}(\rho/\rho_{max} - 0.2, 0)$, where $\rho_{max}$
is the peak density in the torus.  The field is normalized so that the
minimum value of $p_{gas}/p_{mag} = 10^2$.  The orbital period at the
pressure maximum ($r = 12$), is $264$ as measured by an observer at
infinity.

\begin{figure}
\plottwo{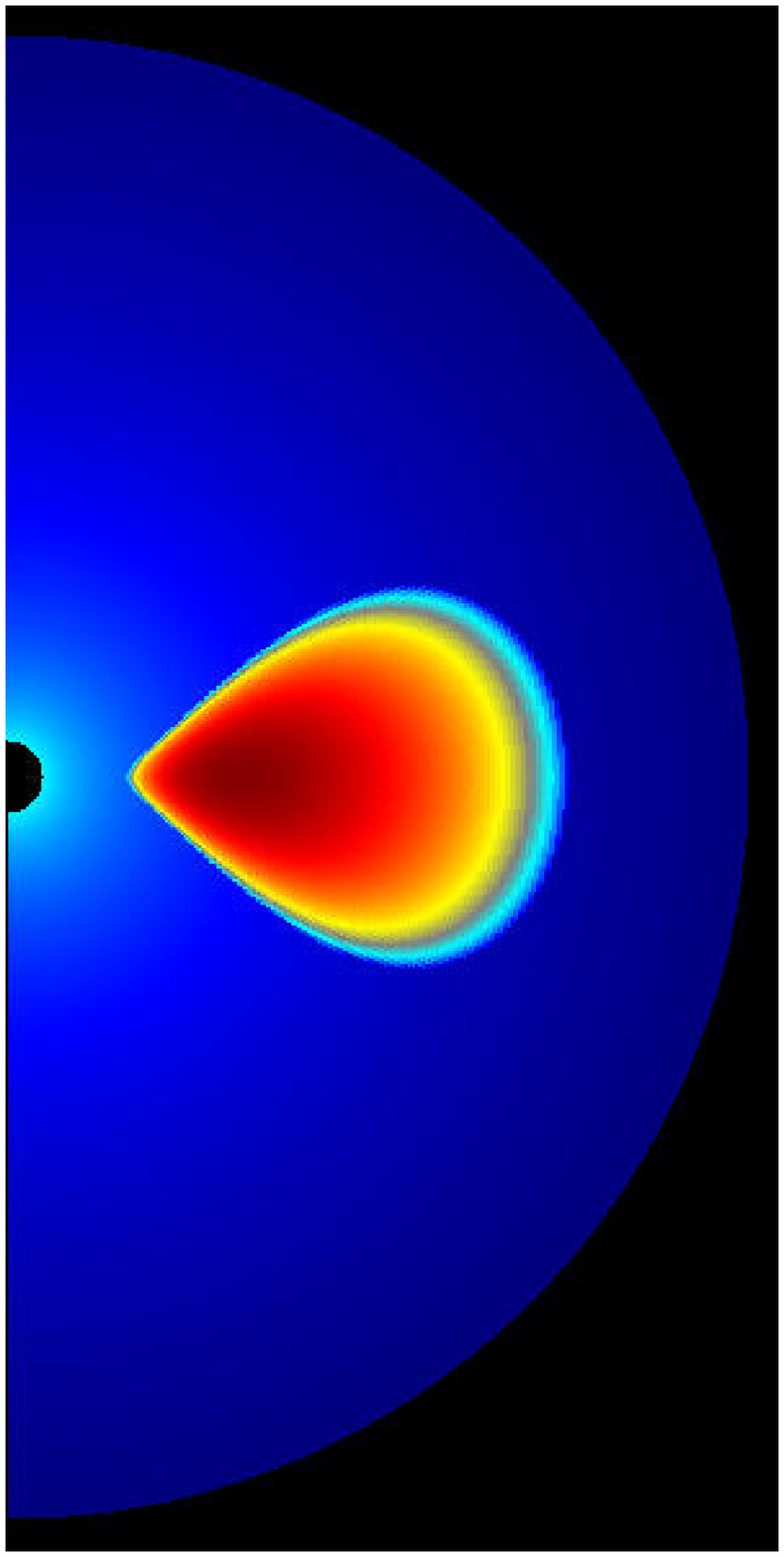}{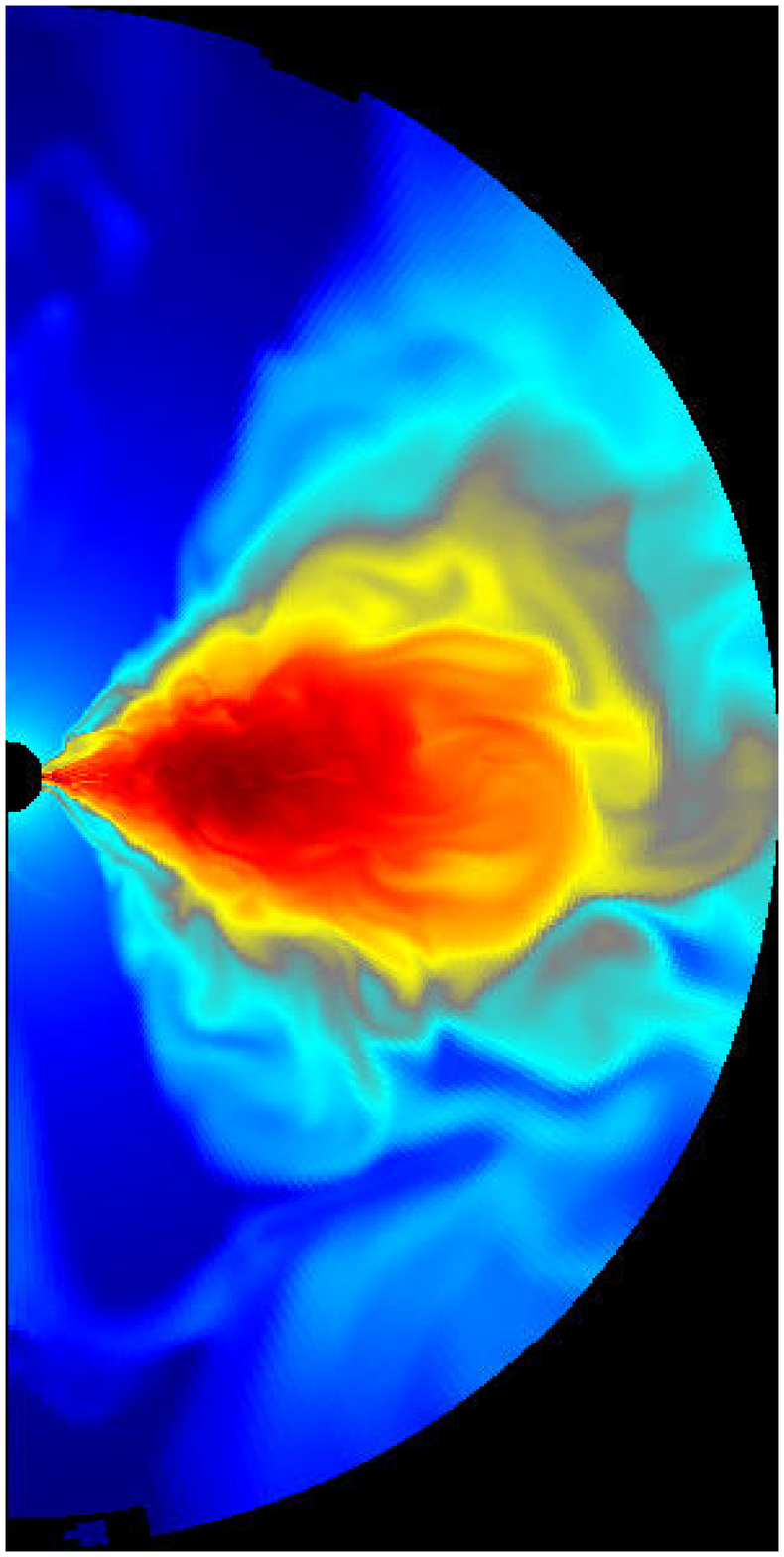}
\caption{
Density field, for a magnetized torus around a Kerr black hole with $a/M
= 0.5$ at $t = 0$ (left) and at $t = 2000 M$ (right).  The color is
mapped from the logarithm of the density; black is low and dark red is
high.  The resolution is $300^2$.
}\label{fig-M}
\end{figure}

The integration extends for $\Delta t = 2000$, or about $7.6$ orbital
periods at the pressure maximum.  Figure \ref{fig-M} shows the initial
and final density states projected on the $R = r \sin(\theta)$, $Z = r
\cos(\theta)$ plane.  Color represents $\log(\rho)$.  The coordinate
parameter $h$, which concentrates zones toward the midplane, is set to
$0.2$.  The torus atmosphere is set to the floor values (see above), and
the MC limiter is used.  The numerical resolution is $300^2$.  

\begin{figure}
\plotone{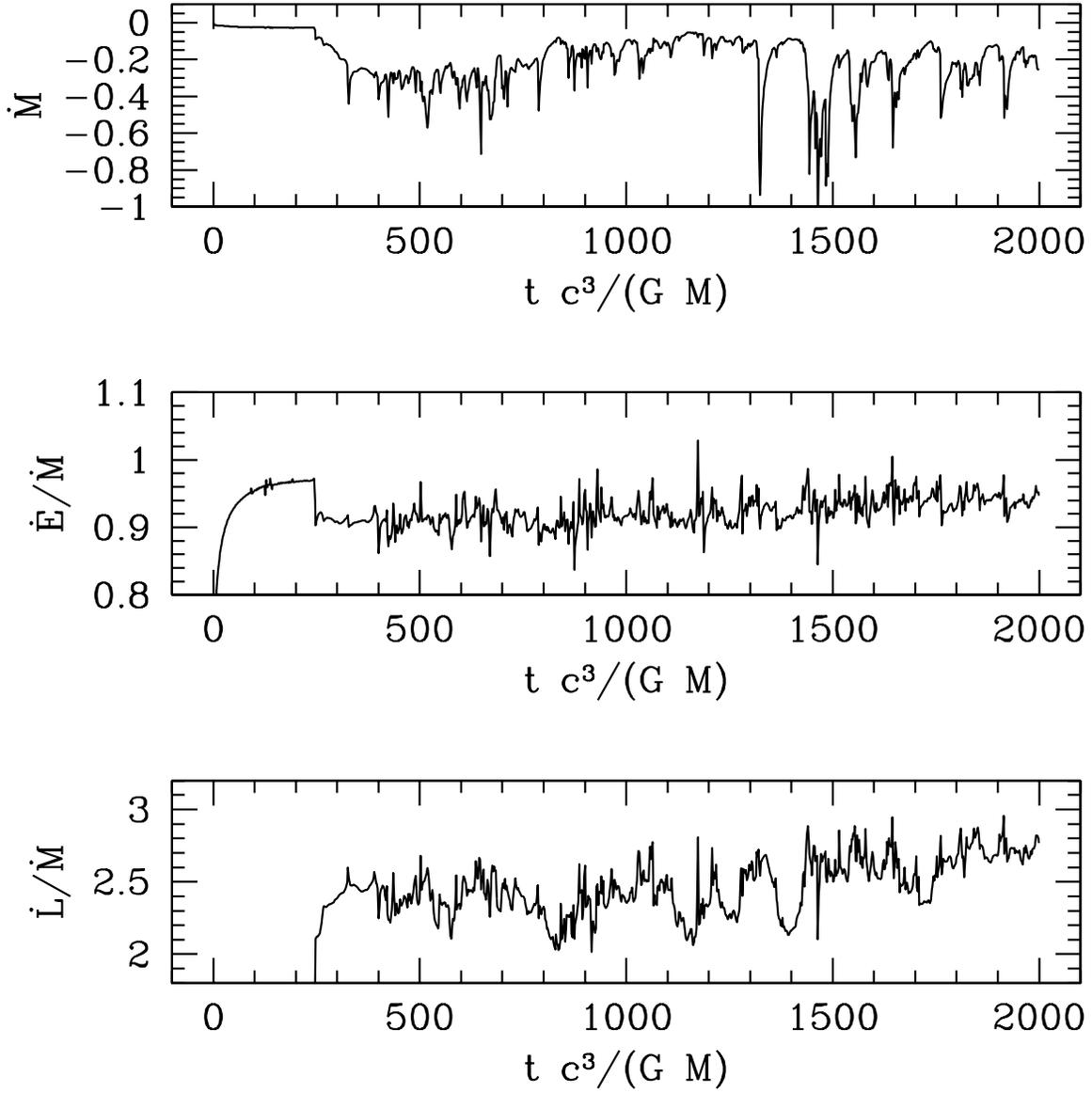}
\caption{
Evolution of the mass accretion rate (top), the specific energy of the
accreted matter (middle), and the specific angular momentum of the
accreted matter (bottom) for a black hole with $a/M = 0.5$.
}\label{fig-O}
\end{figure}

The flux of mass, energy, and angular momentum through the inner
boundary are described in Figure \ref{fig-O}.  Initially the fluxes are
small because the initial conditions are near an (unstable) equilibrium.
The magnetorotational instability \citep{bh91} e-folds for just over an
orbital period, after which the magnetic field has reached sufficient
strength to distort the original torus and drop material into the black
hole.  Later, the torus is turbulent and accretion occurs at a more or
less steady rate.  

\section{Conclusion}

Like all hydrodynamics codes, HARM has failure modes.  We will discuss
one that is likely to be relevant to future astrophysical simulations.
When $B^2/\rho \gg 1$ and $B^2 \gg u$, the magnetic energy is the
dominant term in the total energy equation.  Because the fields are
evolved separately, truncation error in the field evolution can lead to
large fractional errors in the velocity and internal energy.  An example
of this was discussed in \S\ref{MBF}, where the magnetized Bondi flow
test fails for large values of $B^2/\rho$.

Another example can be found in the strong cylindrical explosion problem
of \cite{kom99},  where an overpressured region embedded in a uniform
magnetic field produces a relativistic blast wave.  HARM fails on the
strong-field version of this problem unless we turn the Courant number
down to $0.1$, use the minmod limiter, and sharply increase the accuracy
parameter used in the $\bP(\bU)$ inverter.  This is a particularly
difficult problem, with $B^2/\rho$ as large as $10^4$.  The problems
caused by magnetically dominated regions appears to be generic to
conservative relativistic MHD schemes, where small errors in magnetic
energy density lead to fractionally large errors in other components of
the total energy.  At present this is unavoidable, and has motivated the
development of schemes for the evolution of the electromagnetic field in
the force-free limit \citep{kom02b}.

Finally, to sum up: we have described and tested a code that evolves the
equations of general relativistic magnetohydrodynamics.  This code,
together with the code described in a companion paper by \cite{dvh}, are
the first that stably evolve a relativistic plasma in a Kerr spacetime
for many light crossing times.  The advent of practical, stable GRMHD
codes opens the door for the study of many problems in the theory of
RMRs.  For example, it may be possible to directly evaluate the
importance of magnetic energy extraction from rotating black holes and
the importance of black hole spin in determining jet parameters.  It may
also be possible to couple these schemes to numerical relativity codes
and use them to study dynamical spacetimes with electromagnetic sources.

\acknowledgments

We are grateful to our collaborators John Hawley and Jean-Pierre de
Villiers for extensive discussions.  We are also grateful to Ramesh
Narayan, who supplied the initial inspiration for this project some
years ago.  Stu Shapiro's advice has greatly improved this paper.
Serguei Komissarov and the referee, Luca Del Zanna, made valuable
comments on the manuscript.  This work was supported by NSF ITR grant
PHY 02-05155, by NSF PECASE grant AST 00-93091, by a NASA GSRP
fellowship to JCM, and by a National Center for Supercomputing
Applications (NCSA) faculty fellowship to CFG.  GT is supported by the
Hungarian Science Foundation (OTKA, grant No. T037548) and the Education
Ministry of Hungary (grant No. FKFP-0242-2000).  Some of the numerical
development work for this project was performed at NCSA.

\clearpage

\clearpage

\begin{thebibliography}{}

\bibitem[Abramowicz, Jaroszinski, \& Sikora(1978)]{ajs78} Abramowicz,
	M., Jaroszinski, M., \& Sikora, M. 1978, \aap, 63, 221

\bibitem[Anile(1989)]{anile} Anile, A.M. 1989,  Relativistic Fluids and
	Magneto-fluids, (New York: Cambridge Univ. Press)

\bibitem[Balbus \& Hawley(1991)]{bh91} Balbus, S.A., \& Hawley, J.F.
	1991, \apj, 376, 214

\bibitem[Balbus \& Hawley(1998)]{bh98} Balbus, S.A., \& Hawley, J. H.
	1998, Rev. Mod. Phys., 70, 1

\bibitem[Balsara(2001)]{bal01} Balsara, D. 2001, \apjs, 132, 83

\bibitem[Blandford \& Znajek(1977)]{bz77} Blandford, R.D., \& Znajek, R.
	1977, \mnras, 179, 433

\bibitem[Brio \& Wu(1988)]{bw88} Brio, M., \& Wu, C.C. 1988, JCP, 75,
	500

\bibitem[Del Zanna \& Bucciantini(2002)]{dzb} Del Zanna, L., \&
	Bucciantini, N. 2002, \aap, 390, 1177

\bibitem[Del Zanna et al.(2002)]{dzbl} Del Zanna, L., Bucciantini, N.,
	\& Londrillo, P., 2002, astro-ph/0210618

\bibitem[Evans \& Hawley(1988)]{eh88} Evans, C.R., \& Hawley, J.F. 1988,
	\apj, 332, 659

\bibitem[Fishbone \& Moncrief(1976)]{fm76} Fishbone, L.G., \& Moncrief,
	V. 1976, \apj, 207, 962	

\bibitem[Font(2000)]{font00} Font, J. A. 2000, Liv. Rev. Rel., 3,
	2000-2font

\bibitem[Gammie(1999)]{gam99} Gammie, C.F. 1999, \apjl, 522, L57

\bibitem[Harten et al.(1983)]{hll83} Harten, A., Lax, P.D., \& van Leer,
	B. 1983, SIAM Rev. 25, 35

\bibitem[Hawley, Smarr, \& Wilson(1984)]{hsw} Hawley, J.F., Smarr, L.L.,
	\& Wilson, J.R. 1984, \apjs, 55, 211

\bibitem[Koide et al.(2002)]{kskm} Koide, S., Shibata, K., Kudoh, T.,
	\& Meier, D.L. 2002, Science, 195, 1688

\bibitem[Koide et al.(2000)]{kmsk} Koide, S., Meier, D.L., Shibata, K.,
	\& Kudoh, T. 2000, \apj, VVV, 668

\bibitem[Koide, Shibata, \& Kudoh(1999)]{ksk} Koide, S., Shibata, K., \&
	Kudoh, T. 1999, \apj, 522, 727

\bibitem[Koldoba et al.(2002)]{kku} Koldoba, A.V., Kuznetsov, O.A., 
	\& Ustyugova, G.V.  2002, \mnras, 333, 932

\bibitem[Komissarov(1999)]{kom99} Komissarov, S.S. 1999, \mnras,
	303, 343

\bibitem[Komissarov(2001)]{kom01} Komissarov, S.S. 2001, in
	Godunov Methods: Theory and Applications, ed E.F.Toro, 
	(New York: Kluwer) 519

\bibitem[Komissarov(2002a)]{kom02a} Komissarov, S.S. 2002, astro-ph/0209213

\bibitem[Komissarov(2002b)]{kom02b} Komissarov, S.S. 2002, 
	\mnras, 336, 759

\bibitem[Lax \& Wendroff(1960)]{lw60} Lax, P.D., \& Wendroff, B. 1960,
	Comm. Pure App. Math, 13, 217

\bibitem[LeVeque(1998)]{lev98} LeVeque, R.J. 1998, in Computational
	Methods for Astrophysical Fluid Flow (Berlin: Springer-Verlag)
	1

\bibitem[Marder(1987)]{mar87} Marder, B. 1987, JCP, 68, 48

\bibitem[Mart\'\i~\& Muller(1999)]{mm99} Mart\'\i, J.M., \& M\"uller, E.
	1999, Liv. Rev. in Rel., 2, 1999-3marti

\bibitem[Meier, Koide, \& Uchida(2001)]{mku} Meier, D.L., Koide, S., \&
	Uchida, Y. 2001, Science, 291, 84

\bibitem[Misner, Thorne, \& Wheeler(1973)]{mtw} Misner, C., Thorne, K.,
	\& Wheeler J. 1973, Gravitation, (New York: Freeman) (MTW)

\bibitem[Norman \& Winkler(1986)]{nw86} Norman, M.L., \& Winkler, K.-H.
	1986, in Astrophysical Radiation Hydrodynamics,
	eds. K.-H. Winkler \& M. Norman (Dordrecht: Kluwer), 449

\bibitem[Orszag \& Tang(1979)]{ot79} Orszag, S., \& Tang, C.M. 1979,
	JFM, 90, 129

\bibitem[Phinney(1983)]{ph83} Phinney, E.S., 1983, unpublished Ph.D. thesis,
	Cambridge University

\bibitem[Punsly(2001)]{pun01} Punsly, B. 2000,  Black Hole
	Gravitohydromagnetics, (New York: Springer)

\bibitem[Van Putten(1993)]{van93} Van Putten, M.H.P.M. 1993,
	JCP, 105, 339

\bibitem[Ryu \& Jones(1995)]{rj95} Ryu, D. \& Jones, T.W. 1995, \apj, 442,
	228 (RJ)

\bibitem[Shapiro \& Teukolsky(1983)]{st83} Shapiro, S.L., \& Teukolsky,
	S.A. 1983, Black Holes, White Dwarfs, and Neutron Stars: The Physics of
	Compact Objects (New York: Wiley)

\bibitem[Stone \& Norman(1992)]{sn92} Stone, J.M., \& Norman, M. 1992,
	\apjs, 80, 791

\bibitem[Takahashi et al.(1990)]{tak90} Takahashi, M. , Nitta, S. ,
        Tatematsu, Y.  \& Tomimatsu, A.  1990, \apj, 363, 206

\bibitem[T\'oth(1996)]{toth96} T\'oth, G. 1996, Astrophys. Lett. \&
	Comm. 34, 245

\bibitem[T\'oth(2000)]{toth00} T\'oth, G. 2000, JCP, 161, 605

\bibitem[de Villiers \& Hawley(2002)]{dvh} de Villiers, J.-P., \& Hawley,
	J.F. 2002, \apj, in press

\bibitem[Wilson(1977)]{wil77} Wilson, J.R. 1977, in Proc. of the First
	Marcel Grossman Meeting on General Relativity, ed. R. Ruffini,
	(Amsterdam: North-Holland), 393

\bibitem[Yokosawa(1993)]{yoko93} Yokosawa, M. 1993, PASJ, 45, 207

\end{thebibliography}
\end{document}